\documentclass{iopjournal}
\usepackage[T1]{fontenc}
\usepackage{graphicx} %
\usepackage{amsmath}
\usepackage{booktabs}
\usepackage[capitalise]{cleveref}
\usepackage{float}
\usepackage[caption=false]{subfig}
\usepackage{listings}
\crefname{lstlisting}{Listing}{Listings}
\Crefname{lstlisting}{Listing}{Listings}

\usepackage{tikz}
\usetikzlibrary{arrows.meta}
\usetikzlibrary{decorations.pathmorphing}

\usepackage[sort]{natbib}
\setcitestyle{authoryear, round}

\def\tsc#1{\csdef{#1}{\textsc{\lowercase{#1}}\xspace}}
\tsc{WGM}
\tsc{QE}

\usepackage{acro}
\acsetup{single}
\DeclareAcronym{PET-GEOM}{short=PET-GeoM,long=PET Geometric Efficiency Modelling}
\DeclareAcronym{AFOV}{short=AFOV,long=axial field-of-view}
\DeclareAcronym{LAFOV}{short=LAFOV,long=long axial field-of-view}
\DeclareAcronym{PET}{short=PET,long=Positron Emission Tomography}
\DeclareAcronym{MIRD}{short=MIRD,long=Medical Internal Radiation Dose}
\DeclareAcronym{Ps}{short=Ps,long=positronium}
\DeclareAcronym{o-Ps}{short=o-Ps,long=ortho-positronium}
\DeclareAcronym{p-Ps}{short=p-Ps,long=para-positronium}
\DeclareAcronym{PLI}{short=PLI,long=Positronium Lifetime Imaging}
\DeclareAcronym{TOF}{short=TOF,long=time-of-flight}
\DeclareAcronym{GATE}{short=GATE,long=Geant4 Application for Tomographic Emission}
\DeclareAcronym{MC}{short=MC,long=Monte Carlo}
\DeclareAcronym{LOR}{short=LOR,long=Line of Response}

\newcommand{\blu}[1]{\textcolor{blue}{#1}}

\begin{document}

\articletype{Paper} %

\title{PET-GeoM: Exact determination of two-photon geometric efficiency for arbitrary source positions in cylindrical PET scanners}

\author{Wojciech Zdeb$^{1,*}$\orcid{0009-0000-1485-251X}, Mateusz Bała$^1$\orcid{0000-0002-2505-7949}, Beatrix C. Hiesmayr$^{3,4}$\orcid{0000-0001-9062-6039}, Konrad Klimaszewski$^1$\orcid{0000-0003-0741-5922}, Michał Obara$^1$\orcid{0009-0006-4779-6757}, Lech Raczyński$^1$\orcid{0000-0002-7039-2084}, Roman Y. Shopa$^1$\orcid{0000-0002-1089-5050}, Wojciech Krzemien$^{2,*}$\orcid{0000-0002-9546-358X}}

\affil{$^1$National Centre for Nuclear Research, Department of Complex Systems, 05-400 Otwock, Poland}\\
\affil{$^2$National Centre for Nuclear Research, Department of High Energy Physics, 05-400 Otwock, Poland}\\
\affil{$^3$IT:U Interdisciplinary Transformation University, Freistädter Strasse 400, 4040 Linz, Austria}\\
\affil{$^4$University of Vienna, Faculty of Physics, Währingerstrasse 17, 1090 Vienna, Austria}

\email{$^*$wojciech.krzemien@ncbj.gov.pl, wojciech.zdeb@ncbj.gov.pl}


\begin{abstract}
{\it Objective}.
The sensitivity of a \ac{PET} scanner depends on the source position, yet closed-form results for the two-photon geometric efficiency exist only on the scanner axis, and the off-centre behaviour remains debated: reported NEMA off-centre sensitivities are inconsistent across systems, and a recent analysis suggested positioning patients off-centre to increase count statistics. We aimed to develop an exact, fast and generalisable method for the two-photon geometric efficiency at an arbitrary source position, and to use it to determine how the efficiency depends on the source position and where commonly used approximations hold.

{\it Approach}.
\ac{PET-GEOM} rests on a simple geometric principle: a back-to-back photon pair is detected if and only if its line meets the active detector surface within the overlap of that surface and its central projection through the source. For a cylindrical scanner, the resulting surface integral reduces analytically to a one-dimensional integral. The method, released as an open-source Python library, was validated against a geometric \ac{MC} model, compared with \ac{GATE} simulations and an independent exact model, and applied to point and line sources, and to organ-like regions in four clinical scanner geometries.

{\it Main results}.
\ac{PET-GEOM} agrees with the \ac{MC} reference within its statistical uncertainty and with the independent exact model to $5\times10^{-9}$, at about $50~\mu$s per evaluation. For a point source, the efficiency depends jointly on the transaxial and axial position: in the central plane it decreases with transaxial offset, by 12\% at 10~cm for a scanner with the dimensions of the Philips Vereos, whereas near the end caps it increases. The increase reported for line sources results from axial averaging. For organ-like regions, the sign of the change depends on the fraction of the \ac{AFOV} they cover: at 10~cm, a colon-sized region loses 0.7--3.5\% in the Vision, Quadra and uExplorer geometries but gains 0.95\% in the shorter Vereos. For a line source spanning the \ac{AFOV}, geometry alone predicts an increase of only 0.6--1.7\% at the NEMA 10~cm offset, comparable to detector effects, which may explain the inconsistent reports. Approximations based on the nearer end cap hold only inside two cones spanned by the end-cap rims and the scanner centre, i.e.\ in one-third of the scanner volume.

{\it Significance}.
\ac{PET-GEOM} provides an exact, deterministic and fast reference for the geometric component of \ac{PET} sensitivity in scanner design, sensitivity characterisation and the modelling of sensitivity maps, and shows that a transaxial offset does not in general increase sensitivity. Because the overlap principle holds for any convex detector surface, the approach extends in principle to polygonal rings of flat modules and to detectors with gaps.
\end{abstract}

\section{Introduction}

Sensitivity is one of the key performance metrics of a \ac{PET} scanner. It is commonly approximated as the product of the \textit{geometric efficiency} $g$, describing the fraction of emitted radiation that reaches the detector, and factors describing intrinsic detection efficiency, event selection, attenuation and scatter~\citep{cherry_physics_nuclear_medicine}.
The geometric factor depends only on the scanner geometry and the source position, independently of the detector technology.
Geometric efficiency has become more important with the introduction of \ac{LAFOV} and total-body systems \citep{spencer_performance_2021,badawiFirstHumanImaging2019,vansluisExtendingClinicalCapabilities2022, mehadjiLongAxialField2025,karakatsanisLongAxialFieldofView2026,cookImpactLongAxial2025}. Their gains in sensitivity come primarily from increased solid angle subtended by the detector at the source~\citep{erikssonInvestigationSensitivityLimits2007, baranRealisticTotalbodyJPET2025}.

Computing geometric efficiency for single-particle emission is a standard problem in radiation detection. Analytical and approximate solutions exist for the solid angle subtended by disk and cylindrical detectors at point and extended sources~\citep{conwayGeometricEfficiencyCircular2011,prataAnalyticalCalculationSolid2004, pommeSolidAngleSubtended2007}.
The \ac{PET} case differs considerably, since an annihilation event contributes to the measured data only if both 511 keV, back-to-back photons reach the detector and form a coincidence. The relevant quantity is therefore the fraction of the lines passing through the source that intersect the detector on both sides.
For a source at the centre of a cylindrical scanner of radius $R$ and axial length $2L$, symmetry reduces the problem to a single-photon case. However, this symmetry is lost when the source is displaced from the scanner centre.
The two-photon efficiency then depends on the source position in a way that cannot be inferred from the single-photon solid angle.
This position dependence becomes particularly important in \ac{LAFOV} and total-body systems, where the measured sensitivities vary strongly across the field of view ~\citep{spencer_performance_2021,prenosil_performance_2022}.

This position dependence also matters in practice. Standard characterisation protocols measure scanner sensitivity with a line source at the centre and at 10~cm transaxial offset  (NEMA NU 2-2018~\citep{NEMA:2018}), and reported results are inconsistent: some systems show higher sensitivity off-centre~\citep{chenPerformanceCharacteristicsDigital2020,chicheporticheComparisonNEMACharacterizations2020}, others the opposite~\citep{jakobyPhysicalClinicalPerformance2011}. Recently, \citet{Lin_2026} derived a solid-angle model of transaxial sensitivity. They concluded that sensitivity increases with transaxial offset and peaks at the edge of the field of view, and suggested that patients could be offset deliberately to increase count statistics. Their analysis, however, considered an axially extended line source, whose response is an average over all axial positions. How the two-photon efficiency of a point source depends jointly on its transaxial and axial position, and therefore which of these trends are general properties of cylindrical geometry, has not been examined systematically.

An exact and fast description of the $g$ factor as a function of source position is useful in several contexts: First, for comparative studies of scanner geometries at the design stage. Second, for estimating line-source sensitivity as part of the standard procedure for NEMA scanner characterisation. Third, as a baseline reference in studies of sensitivity maps and geometric normalisation components for image reconstruction~\citep{iriarteSystemModelsPET2016}.
Existing tools do not fully meet this need. Closed-form expressions are established only for sources on the scanner axis ~\citep{Bonifacio2023}. Simple approximations based on the angular acceptance of the nearer end of the scanner are used for off-axis positions, but their domain of validity has not been established.
Monte Carlo toolkits such as \acf{GATE}~\citep{sarrutAdvancedMonteCarlo2021a,sarrutGATE10Monte2026,krahGATE10Monte2026} provide position-dependent sensitivity but require a separate simulation for every geometry and source position. Moreover, isolating the purely geometric component from such simulations requires additional work to separate it from detector and physics effects.

In this work, we address these issues with the following contributions. First, we derive an exact, self-contained formulation of the two-photon geometric efficiency of a cylindrical detector for a point source at an arbitrary position.
The method is based on a geometric construction: the set of admissible intersection points on the detector is the overlap between the cylinder and its projection through the source point.
Next, using this formulation, we map the efficiency over the full scanner volume. We show that its dependence
on the transaxial offset is non-monotonic and changes trend with axial position, and that the increase reported for line sources results from axial averaging combined with the two-photon condition.
Also, we identify the domains of applicability of the two approximate models. We show that simple approximations are valid only inside two cones defined by the end-cap and the scanner centre, and fail in the remaining two-thirds of the scanner volume. We provide a geometric explanation of this boundary.
Finally, we provide an open-source, deterministic implementation \acf{PET-GEOM}, which evaluates $g$ factor
at a fraction of the cost of Monte Carlo sampling.

We validate \ac{PET-GEOM} against a purely geometric Monte Carlo simulation as an independent reference, and compare it qualitatively with \ac{GATE} simulations including detector response. We compare it with two approximate models and with the corrected model of~\citet{Lin_2026}, with which it agrees.  The comparison covers: (i) setups with point sources over a wide range of axial and transaxial offsets, (ii) a line source (thin rod) study reproducing the configuration of~\citet{Lin_2026}, and (iii) organ-like regions offset transaxially in four clinical scanner geometries.

The paper is organised as follows. \Cref{section:notation} introduces the notation used throughout the paper, including the cylindrical coordinate system and common variables. \Cref{section:sens_models} presents the theoretical foundations of the sensitivity models considered in the comparison. \Cref{section:implementation} describes the software and numerical implementation of the proposed model. \Cref{section:mc_simulations} presents the \ac{MC} simulations used as a reference for evaluating the accuracy of the models under consideration. \Cref{section:results} presents the results of the comparison, followed by a discussion of the findings in \cref{section:discussion}. Finally, \cref{section:conclusions} summarises the main conclusions of the study.

\section{Geometry and definitions}
\label{section:notation}
\subsection{Detector model and coordinates}
\begin{figure}[htbp]
    \centering
    \begin{tikzpicture}
    \pgfmathsetmacro{\L}{1.3*8}
    \pgfmathsetmacro{\R}{1.3*2.5}
    \pgfmathsetmacro{\zZero}{\L/4}
    \pgfmathsetmacro{\rad}{3*\R/5}

    \draw[dashed, arrows=-Latex] (-\L/2 - 1.5,0) -- (\L/2 + 1.5,0);
    \draw[dashed, arrows=-Latex] (0,-7*\R/5) -- (0,7*\R/5);

    \node[draw=none, shape=circle] at (\L/2 + 1.6, 0.17) {z};
    \node[draw=none, shape=circle] at (0.17, 7*\R/5 + 0.2) {$\rho$};

    \draw[dashed, line width=1.5pt] (\L/2,\R) arc [start angle=90, end angle=270,
         x radius=0.5, y radius=\R];

    \draw[
        color=green!90,
        line width=1pt,
        decorate,
        decoration={snake, amplitude=2pt, segment length=8pt, post length=7pt, pre length=7pt},
        Latex-Latex,
    ] (3*\L/8,\R) -- (-\L/4,-\R);

    \draw[
        color=red!90,
        line width=1pt,
        decorate,
        decoration={snake, amplitude=2pt, segment length=8pt, post length=7pt, pre length=7pt},
        Latex-Latex,
    ] (\L/16,\R) -- (11*\L/16,-5*\R/16);

    \draw[line width=2pt] (-\L/2,0) ellipse (0.5 and \R);
    \draw[line width=2pt] (\L/2,-\R) arc [start angle=-90, end angle=90,
         x radius=0.5, y radius=\R];

    \draw[line width=2pt] (-\L/2,\R) -- (\L/2,\R);
    \draw[line width=2pt] (-\L/2,-\R) -- (\L/2,-\R);

    \node (source) at (0,0) [circle, fill=black!100,minimum size=0.5pt,label=60:$O$] {};
    \node (source) at (\zZero,\rad) [circle, fill=black!100,minimum size=0.5pt,label=0:$S$] {};

    \draw[dashed, line width=1.5pt, darkgray]
        (-\L/2,-\R-0.4) -- (\L/2,-\R-0.4);
    \node at (0.5,-\R-0.85) {$2L$};
    \draw[line width=1.5pt, darkgray] (-\L/2,-\R-0.1) -- (-\L/2,-\R-0.8);
    \draw[line width=1.5pt, darkgray] (\L/2,-\R-0.1) -- (\L/2,-\R-0.8);

    \draw[dash dot,line width=1pt]
        (-\L/2,0) -- (-\L/2,\R)
        node[midway, left=0.7pt] {$R$};

    \draw[dash dot,line width=1pt] (\zZero, 0) -- (\zZero,\R);

    \draw[dashed,line width=1.5pt, darkgray]
        (0,\R+0.4) -- (\zZero,\R+0.4)
        node[midway, above=2pt] {$z_S$};
    \draw[line width=1.5pt, darkgray] (\zZero,\R+0.1) -- (\zZero,\R+0.8);

    \draw[dash dot,line width=0.9pt] (0, \rad) -- (\zZero,\rad);

    \draw[line width=1.5pt] (-0.15, \rad) -- (0.15, \rad)
        node[midway, left=1.3pt] {$\rho_S$};

\end{tikzpicture}
    \caption{Idealised cylindrical detector geometry, coordinate system, and source position $S=(\rho_S, 0,z_S)$, where $\phi_S=0$ is chosen without loss of generality due to the rotational symmetry of the detector. The green and red arrows illustrate photon emission directions that do and do not permit the detection of a coincidence event, respectively.}
    \label{fig:detector_geometry}
\end{figure}
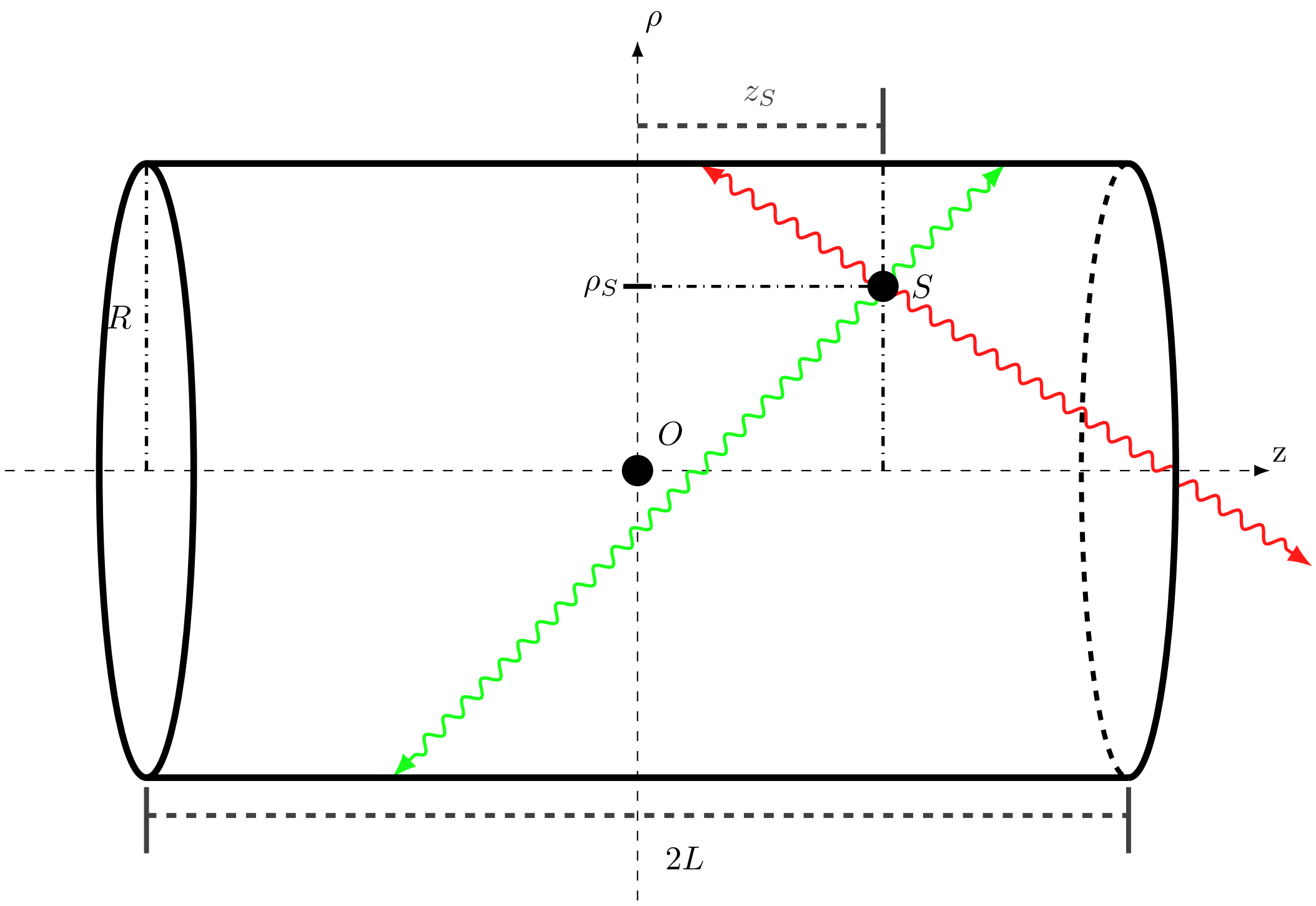

Throughout this work, a cylindrical coordinate system $(\rho,\phi,z)$ is used, with the $z$-axis coinciding with the symmetry axis of the cylindrical detector and the origin $O$ at the scanner centre.
The radial coordinate $\rho$ denotes the distance from the $z$-axis, while $\phi$ is the azimuthal angle around this axis.
Cylindrical \ac{PET} detector systems are approximated by a complete cylindrical surface with inner radius $R$ and total axial length $2L$. Thus, the detector volume is defined as the cylindrical region $\rho\leq R$, $-L\leq z\leq L$,
whereas the active detector surface is restricted to the lateral surface $\rho=R$, $-L\leq z\leq L$.
The cylindrical end caps at $z=\pm L$ are not included in the detector surface.
We denote the lateral surface of the cylinder as $C$, and the two open end caps as $D_1$ and $D_2$, respectively.
The use of a cylindrical surface provides an idealised representation of cylindrical detector systems while retaining the geometric characteristics relevant to the calculation of geometric efficiency. The detector geometry and the coordinate system are shown in \cref{fig:detector_geometry}.

The position of a point source is given in cylindrical coordinates as $S=(\rho_S,\phi_S,z_S)$, where $\rho_S$, $\phi_S$, and $z_S$ denote the radial, azimuthal, and axial source coordinates, respectively. For calculations performed in Cartesian coordinates, the same position is expressed as
\begin{equation}
    S=(x_S,\,y_S,\,z_S)=(\rho_S\cos\phi_S,\;\rho_S\sin\phi_S,\;z_S).
    \label{eq:S_cart}
\end{equation}

By rotational symmetry we set $\phi_S=0$, and by the reflection
symmetry $z\to -z$ we may take $z_S\ge 0$ without loss of generality.
We denote
by $D_1$ the end cap nearer to the source, and define the axial distances from
the source to the near and far end caps,
\begin{equation}
  a = L-z_S, \qquad b = L+z_S .
\end{equation}

The source emits isotropically pairs of exactly collinear, back-to-back photons.
Positron range, photon acollinearity, and attenuation are neglected.

\subsection{Single- and two-photon geometric efficiency}
    \label{subsection:single_vs_two_geom_eff}

Two distinct quantities are commonly referred to as geometric efficiency.
Let $\Omega(\cdot)$ denote the solid angle subtended at the source position.
For a source inside the detector volume, surfaces $C, D_1$ and $D_2$ enclose the source and therefore:
\begin{equation}
    \Omega(C) +\Omega(D_1)+\Omega(D_2)  =4 \pi
\end{equation}
Since the sensitive part of the detector corresponds to the lateral surfaces,
the single photon efficiency can be defined as the fraction of the solid angle of the lateral sides:
\begin{equation}
    g_{1\gamma} = \frac{\Omega(C)}{4\pi}
\end{equation}
or equivalently by the complement:
\begin{equation}
    g_{1\gamma} = 1 -\frac{\Omega(D_1) + \Omega(D_2)}{4\pi}
\end{equation}
since a single photon can escape through either end.

For two photons emitted back-to-back, the relevant quantity is the probability that both photons intersect the lateral surface of the detector. Unlike the single-photon case, the geometric acceptance of a back-to-back photon pair is determined by a single \ac{LOR} passing through the source, with the two photons propagating in opposite directions along that line.

The emission directions can be classified into three mutually exclusive categories: neither photon intersects the lateral surface, exactly one photon intersects it, or both photons intersect it. Let $\Omega_0$, $\Omega_1$, and $\Omega_2$ denote the corresponding solid angles in the space of emission directions. These categories exhaust all possible directions, so that \begin{equation}
    \Omega_0 + \Omega_1 + \Omega_2 = 4\pi.
\end{equation}
The two-photon geometric efficiency can therefore be defined directly as the fraction of directions for which both photons intersect the lateral surface:
\begin{equation}
    g_{2\gamma} = \frac{\Omega_2}{4\pi},
\end{equation}
or, equivalently, by excluding the two classes of rejected photon pairs:
\begin{equation}
    g_{2\gamma} = 1 - \frac{\Omega_0 + \Omega_1}{4\pi}.
\end{equation}
The latter expression makes explicit that a coincidence is rejected whenever at least one photon fails to intersect the lateral surface. The two-photon efficiency thus depends on the joint geometric acceptance of the pair, rather than on the acceptance of either photon considered independently.
The two expressions are mathematically equivalent, but they lead to different computational routes: \ac{PET-GEOM} evaluates $\Omega_2$ directly (\cref{subsection:pet_geom}), whereas the model of \citet{Lin_2026} evaluates the rejected fraction through the end caps (\cref{subsection:lin_et_al_model}). Comparing the two methods therefore also tests this equivalence numerically (\cref{subsection:validation}).

The $g_{1\gamma}$  and $g_{2\gamma}$ coincide for the source placed in the geometric centre of the detector. Otherwise, their behaviour is rather different.
In the rest of the article, we consider mostly the two-photon geometric efficiency and for simplicity we drop the $2\gamma$ index, so: $g \equiv g_{2\gamma}$.

\section{Sensitivity Models}\label{section:sens_models}
The geometric efficiency of a \ac{PET} detector for two-gamma, back-to-back photon emission depends on the relative position of the source and the detector geometry. Several approaches can be used to estimate this quantity, differing in the mathematical formulation and in the approximations required to obtain the final result.

In this work, four methods are considered. Two are analytical approximations: the first is based on a Taylor expansion of an expression for a source displaced from the detector centre, while the second estimates the geometric efficiency from the fractions of solid angles subtended by the detector for the two possible photon directions. The remaining two methods provide accurate results, but require numerical integration, including the model developed in this work.

The geometric model developed in this work is presented first, followed by a brief introduction to previously proposed approaches used for comparison. Particular attention is devoted to the derivation of the proposed model and to the assumptions underlying its formulation, as these assumptions determine its applicability to different source positions and detector geometries.

\subsection{The \ac{PET-GEOM} method}
\label{subsection:pet_geom}

\subsubsection{Admissible region of the detector}

Assume that we have a cylindrical \ac{PET} detector, with radius $R$ and \acf{AFOV} $2L$. We can approximate the detector's shape as a hollow cylinder with both caps removed. In Cartesian coordinates, a single point on the surface of such a solid is then parametrised as:

\begin{equation}
    p = (R\cdot\cos{\phi}, R\cdot\sin{\phi}, z)
    \label{eq:p_param}
\end{equation}
where $\phi\in [0, 2\pi]$ and $z\in [-L, L]$.

Next, consider a point source, given as $S = (\rho_S, \phi_S, z_S )$, and located within the volume of said cylindrical detector. From $S$, a pair of photons is emitted back-to-back, such that their direction vectors are antiparallel. Consequently, the two-photon emission can be represented geometrically by a straight line $l$ passing through the source point $S$, with the two photons propagating in opposite directions along $l$.
To determine the geometric efficiency for the given source position, we need to identify the emission directions for which the corresponding line $l$ intersects the finite cylindrical detector at two distinct points. Equivalently, for each possible first intersection point on the detector surface, we ask whether the same line intersects the cylindrical surface again before leaving the finite axial extent of the detector.

To formalise this question, we first consider the set of all possible locations of the first intersection point, i.e. the lateral surface of the finite cylinder:

\begin{equation}
    C = \left\{
\left(R\cos\phi,\; R\sin\phi,\; z\right)
\;\middle|\;
\phi \in [0,2\pi],\;
z \in \left[-L,L\right]
\right\}
\end{equation}

For determining whether a second intersection exists, it is convenient to temporarily disregard the axial boundaries of the detector and consider an \textit{infinitely long} cylinder of the same radius:

\begin{equation}
    C_{\infty} = \left\{
\left(R\cos\phi,\; R\sin\phi,\; z\right)
\right\}
\end{equation}

Since the interval $\phi \in [0,2\pi)$ already spans the entire circumference of the cylinder, any value of $\phi$ outside this interval can be mapped to an equivalent value within it. Therefore, removing the bounds on $\phi$ does not change the geometric region under consideration, leaving an infinitely long "tube" of radius $R$. In essence, when we are looking for intersections with this infinitely long cylinder, we are projecting each point $p\in C$ onto $C_{\infty}$, such that the original point $p$ and its projection $p'$ are collinear with each other and with the point $S$.

Now, what we will find is that the collection of points $p'$ will form a new solid which is "carved out" of $C_{\infty}$, and also that both the original cylinder $C$ and this new solid will "overlap" in some capacity, as long as the source point $S$ lies within the volume of the detector. If we then define the image of $C$ onto the infinite cylinder $C_{\infty}$ as $C'$, then the answer to the question asked a few paragraphs earlier is that the first intersection point has to necessarily lie within that overlap, i.e.:

\begin{equation}
    \forall p_{int}: p_{int}\in C \cap C'
\end{equation}

This holds true due to the presented process being reversible and, consequently, all of the conclusions drawn from this process. As such, instead of considering $C$ as our original cylinder, we could consider this new shape $C'$ to be our starting solid, and since the operation of central projection is reversible, giving us $(p')' = p$, then necessarily $(C')' = C$.
From that, it is easy to see that if the first intersection were to lie within a non-overlapping part of $C'$, i.e. $p_{int}\in C' \setminus C$, then that would already disqualify it since such a point necessarily does not intersect $C$, which violates our conditions. Similarly, we can discard all such points that $p_{int}\in C \setminus C'$ simply by reversing the logic, finally arriving at the postulated $p_{int}\in C \cap C'$.

The intersection region $C\cap C'$ therefore comprises precisely those points on the lateral surface that can serve as the first intersection point of a photon trajectory for which a second intersection with the detector also exists. Equivalently, this region identifies the accepted emission directions for which both photons intersect the lateral surface. Its solid angle consequently corresponds to $\Omega_2$, as defined in \cref{subsection:single_vs_two_geom_eff}, and the two-photon geometric efficiency can be expressed as
\begin{equation}
    g=\frac{\Omega(C\cap C')}{4\pi} = \frac{\Omega_2}{4\pi}.
    \label{eq:g_overlap}
\end{equation}

\subsubsection{Surface integral}
We define the intersection region $C\cap C'$ as follows. First, due to cylindrical symmetry, $\phi_S = 0$ is chosen without loss of generality. The source can therefore be represented in the $\phi=0$ cross-section, which we identify with the Cartesian $x$-$z$ plane. In this plane, the radial coordinate $\rho$ coincides with the Cartesian coordinate $x$, so that

\begin{equation}
    S \rightarrow \left(\rho_S, 0, z_S\right)
\end{equation}
This choice does not affect the result, while significantly simplifying the calculations. Consequently, the projected geometry can be determined by evaluating the mappings of the following four points expressed in the Cartesian $x$-$z$ plane:

\begin{equation}
    \left(R, 0, \pm L\right),
    \qquad
    \left(-R, 0, \pm L\right).
\end{equation}

\begin{figure}[htbp]
    \centering
    \subfloat[\label{fig:diagramA_corner_projections}]{%
        \resizebox{0.7\textwidth}{!}{%
            \begin{tikzpicture}
    \pgfmathsetmacro{\L}{7}
    \pgfmathsetmacro{\R}{2.5}
    \pgfmathsetmacro{\zZero}{\L/16}
    \pgfmathsetmacro{\rad}{\R/4}

    \draw[thin, dashed, arrows=-Latex] (-\L/2 - 1.5,0) -- (\L + 0.5,0);
    \draw[thin, dashed, arrows=-Latex] (0,-6*\R/5) -- (0,6*\R/5);

    \node[draw=none, shape=circle] at (\L + 0.6, 0.17) {z};
    \node[draw=none, shape=circle] at (0.17, 6*\R/5 + 0.2) {$x$};

    \draw[line width=1.2pt] (-\L/2,-\R) rectangle (\L/2,\R);

    \node[above left]  at (-\L/2,\R)  {$p_1$};
    \node[above right] at (\L/2,\R)   {$p_2$};
    \node[below right] at (\L/2,-\R)  {$p_3$};
    \node[below left]  at (-\L/2,-\R) {$p_4$};

    \draw[line width=1.2pt,dash dot,gray] (-\L/2 - 1.5,-\R) -- (-\L/2,-\R);
    \draw[line width=1.2pt,dash dot,gray] (\L/2,-\R) -- (\L + 0.5,-\R);
    \draw[line width=1.2pt,dash dot,gray] (-\L/2 - 1.5,\R) -- (-\L/2,\R);
    \draw[line width=1.2pt,dash dot,gray] (\L/2,\R) -- (\L + 0.5,\R);

    \fill[black] (\zZero,\rad) circle (3pt)
        node[above] {$S$};

    \pgfmathsetmacro{\xUL}{\zZero + (-\L/2-\zZero)/(\R-\rad)*(-\R-\rad)}
    \pgfmathsetmacro{\xLL}{\zZero + (-\L/2-\zZero)/(-\R-\rad)*(\R-\rad)}
    \pgfmathsetmacro{\xUR}{\zZero + (\L/2-\zZero)/(\R-\rad)*(-\R-\rad)}
    \pgfmathsetmacro{\xLR}{\zZero + (\L/2-\zZero)/(-\R-\rad)*(\R-\rad)}

    \draw[blue,dashed] (-\L/2,\R) -- (\xUL,-\R);
    \draw[blue,dashed] (-\L/2,-\R) -- (\xLL,\R);
    \draw[blue,dashed] (\L/2,\R) -- (\xUR,-\R);
    \draw[blue,dashed] (\L/2,-\R) -- (\xLR,\R);

    \fill[blue] (\xUL,-\R) circle (3pt);
    \fill[blue] (\xLL,\R) circle (3pt);
    \fill[blue] (\xUR,-\R) circle (3pt);
    \fill[blue] (\xLR,\R) circle (3pt);

    \node[below left]  at (\xUL,-\R) {$p'_1$};
    \node[below left] at (\xUR,-\R) {$p'_2$};
    \node[above right] at (\xLR,\R)  {$p'_3$};
    \node[above left]  at (\xLL,\R)  {$p'_4$};
\end{tikzpicture}
        }%
    }
    \vspace{0.5em}
    \subfloat[\label{fig:diagramB_intervals}]{%
        \resizebox{0.7\textwidth}{!}{%
            \begin{tikzpicture}
    \pgfmathsetmacro{\L}{7}
    \pgfmathsetmacro{\R}{2.5}
    \pgfmathsetmacro{\zZero}{\L/16}
    \pgfmathsetmacro{\rad}{\R/4}

    \draw[thin, dashed, arrows=-Latex] (-\L/2 - 1.5,0) -- (\L + 0.5,0);
    \draw[thin, dashed, arrows=-Latex] (0,-6*\R/5) -- (0,7*\R/5);

    \node[draw=none, shape=circle] at (\L + 0.6, 0.17) {z};
    \node[draw=none, shape=circle] at (0.17, 7*\R/5 + 0.2) {$x$};

    \draw[line width=1.2pt] (-\L/2,-\R) rectangle (\L/2,\R);

    \draw[line width=1.2pt,dash dot,gray] (-\L/2 - 1.5,-\R) -- (-\L/2,-\R);
    \draw[line width=1.2pt,dash dot,gray] (\L/2,-\R) -- (\L + 0.5,-\R);
    \draw[line width=1.2pt,dash dot,gray] (-\L/2 - 1.5,\R) -- (-\L/2,\R);
    \draw[line width=1.2pt,dash dot,gray] (\L/2,\R) -- (\L + 0.5,\R);

    \fill[black] (\zZero,\rad) circle (3pt)
        node[above] {$S$};

    \pgfmathsetmacro{\xUL}{\zZero + (-\L/2-\zZero)/(\R-\rad)*(-\R-\rad)}
    \pgfmathsetmacro{\xLL}{\zZero + (-\L/2-\zZero)/(-\R-\rad)*(\R-\rad)}
    \pgfmathsetmacro{\xUR}{\zZero + (\L/2-\zZero)/(\R-\rad)*(-\R-\rad)}
    \pgfmathsetmacro{\xLR}{\zZero + (\L/2-\zZero)/(-\R-\rad)*(\R-\rad)}

    \draw[blue,dashed] (-\L/2,\R) -- (\xUL,-\R);
    \draw[blue,dashed] (-\L/2,-\R) -- (\xLL,\R);
    \draw[blue,dashed] (\L/2,\R) -- (\xUR,-\R);
    \draw[blue,dashed] (\L/2,-\R) -- (\xLR,\R);

    \draw[blue, line width=1.4pt] (\xUL,-\R) -- (\xLL,\R);
    \draw[blue, line width=1.4pt] (\xLL,\R) -- (\xLR,\R);
    \draw[blue, line width=1.4pt] (\xLR,\R) -- (\xUR,-\R);
    \draw[blue, line width=1.4pt] (\xUR,-\R) -- (\xUL,-\R);

    \fill[blue] (\xUL,-\R) circle (3pt);
    \fill[blue] (\xLL,\R) circle (3pt);
    \fill[blue] (\xUR,-\R) circle (3pt);
    \fill[blue] (\xLR,\R) circle (3pt);

    \node[above right]  at (\xUL,-\R) {$p'_1$};
    \node[above left] at (\xUR,-\R) {$p'_2$};
    \node[below right] at (\xLR,\R)  {$p'_3$};
    \node[below left]  at (\xLL,\R)  {$p'_4$};

    \draw[thick, dashed] (\xUR,-\R-0.3) -- (\xUL,-\R-0.3);
    \draw[thick] (\xUR,-\R-0.45) -- (\xUR,-\R-0.15);
    \draw[thick] (\xUL,-\R-0.45) -- (\xUL,-\R-0.15);
    \node[below] at ({(\xUR+\xUL)/2},-\R-0.45) {$z_{\mathrm{bottom}}$};

    \draw[thick, dashed] (\xLL,\R+0.3) -- (\xLR,\R+0.3);
    \draw[thick] (\xLL,\R+0.15) -- (\xLL,\R+0.45);
    \draw[thick] (\xLR,\R+0.15) -- (\xLR,\R+0.45);
    \node[above] at ({(\xLL+\xLR)/2},\R+0.45) {$z_{\mathrm{top}}$};
\end{tikzpicture}
        }%
    }
    \vspace{0.5em}
    \subfloat[\label{fig:diagramC_overlap}]{%
        \resizebox{0.7\textwidth}{!}{%
            \begin{tikzpicture}
    \pgfmathsetmacro{\L}{7}
    \pgfmathsetmacro{\R}{2.5}
    \pgfmathsetmacro{\zZero}{\L/16}
    \pgfmathsetmacro{\rad}{\R/4}

    \draw[thin, dashed, arrows=-Latex] (-\L/2 - 1.5,0) -- (\L + 0.5,0);
    \draw[thin, dashed, arrows=-Latex] (0,-6*\R/5) -- (0,6*\R/5);

    \node[draw=none, shape=circle] at (\L + 0.6, 0.17) {z};
    \node[draw=none, shape=circle] at (0.17, 6*\R/5 + 0.2) {$x$};

    \draw[line width=1.2pt] (-\L/2,-\R) rectangle (\L/2,\R);

    \node[above left]  at (-\L/2,\R)  {$p_1$};
    \node[above right] at (\L/2,\R)   {$p_2$};
    \node[below right] at (\L/2,-\R)  {$p_3$};
    \node[below left]  at (-\L/2,-\R) {$p_4$};

    \draw[line width=1.2pt,dash dot,gray] (-\L/2 - 1.5,-\R) -- (-\L/2,-\R);
    \draw[line width=1.2pt,dash dot,gray] (\L/2,-\R) -- (\L + 0.5,-\R);
    \draw[line width=1.2pt,dash dot,gray] (-\L/2 - 1.5,\R) -- (-\L/2,\R);
    \draw[line width=1.2pt,dash dot,gray] (\L/2,\R) -- (\L + 0.5,\R);

    \fill[black] (\zZero,\rad) circle (3pt)
        node[above] {$S$};

    \pgfmathsetmacro{\xUL}{\zZero + (-\L/2-\zZero)/(\R-\rad)*(-\R-\rad)}
    \pgfmathsetmacro{\xLL}{\zZero + (-\L/2-\zZero)/(-\R-\rad)*(\R-\rad)}
    \pgfmathsetmacro{\xUR}{\zZero + (\L/2-\zZero)/(\R-\rad)*(-\R-\rad)}
    \pgfmathsetmacro{\xLR}{\zZero + (\L/2-\zZero)/(-\R-\rad)*(\R-\rad)}

    \draw[blue, line width=1.4pt] (\xUL,-\R) -- (\xLL,\R);
    \draw[blue, line width=1.4pt] (\xLL,\R) -- (\xLR,\R);
    \draw[blue, line width=1.4pt] (\xLR,\R) -- (\xUR,-\R);
    \draw[blue, line width=1.4pt] (\xUR,-\R) -- (\xUL,-\R);

    \begin{scope}
        \clip (-\L/2,-\R) rectangle (\L/2,\R);

        \fill[
            blue,
            fill opacity=0.25
        ]
        (\xUL,-\R) --
        (\xLL,\R) --
        (\xLR,\R) --
        (\xUR,-\R) --
        cycle;
    \end{scope}

    \fill[blue] (\xUL,-\R) circle (3pt);
    \fill[blue] (\xLL,\R) circle (3pt);
    \fill[blue] (\xUR,-\R) circle (3pt);
    \fill[blue] (\xLR,\R) circle (3pt);

    \node[below left]  at (\xUL,-\R) {$p'_1$};
    \node[below left] at (\xUR,-\R) {$p'_2$};
    \node[above right] at (\xLR,\R)  {$p'_3$};
    \node[above left]  at (\xLL,\R)  {$p'_4$};
\end{tikzpicture}
        }%
    }
    \caption{Overview of the \ac{PET-GEOM} method: (a) Projections of the cylinder corners onto $C_{\infty}$ along lines passing through the source point $S$, defining the projected corner positions $p_i'$. (b) Axial intervals bounded by the projected corner positions, illustrating the cross-sections associated with the original cylindrical detector and its projection. (c) Overlap of the projected regions, shown in blue, which identifies the admissible region used to determine the two-photon geometric efficiency.}
    \label{fig:petgem_method}
\end{figure}
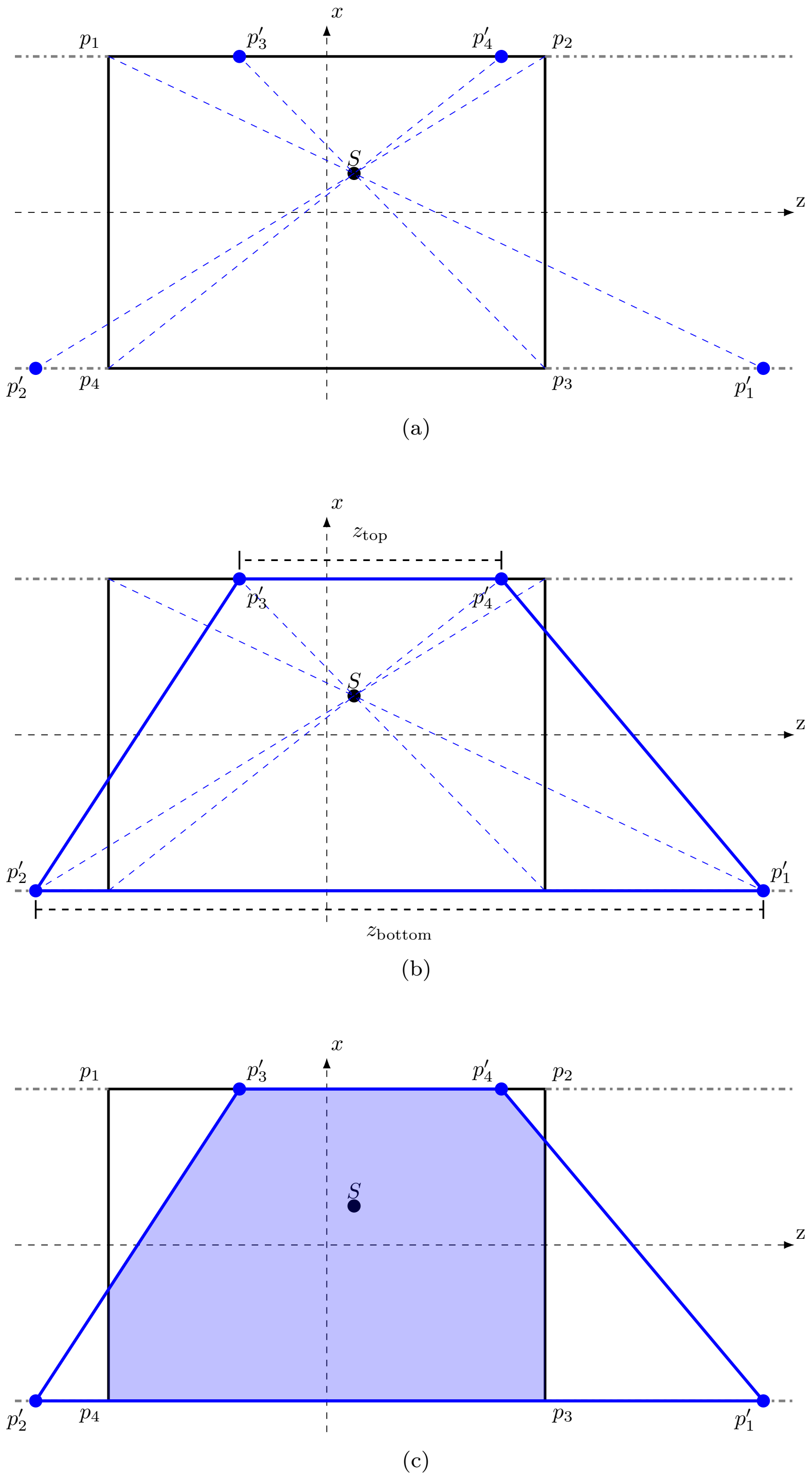

For a boundary point $p=(x_p, y_p, z_p)$, the mapped axial coordinate is computed from the collinearity condition between the source position $S$, the original detector point $p$, and its mapped counterpart $p'$ (\cref{fig:diagramA_corner_projections}):

\begin{equation}
    z_{p'} =
    z_S
    +
    \frac{
        \left(-x_p - x_S\right)\left(z_p-z_S\right)
    }{
        x_p-x_S
    }.
\end{equation}
Evaluating this expression for the four detector corners yields two projected axial intervals (\cref{fig:diagramB_intervals}):
\begin{equation}
    z_{\mathrm{bottom}}
    =
    \left(
    z_{p'_2},
    z_{p'_1}
    \right),
\end{equation}
and
\begin{equation}
    z_{\mathrm{top}}
    =
    \left(
    z_{p'_3},
    z_{p'_4}
    \right).
\end{equation}
These projected points define the boundary of the overlap region $C\cap C'$ in the $x$-$z$ cross-section.

The overlap contour (\cref{fig:diagramC_overlap}) consists of three separate regions. In the central region, the detector remains fully visible, and the angular acceptance spans the complete half-cylinder:

\begin{equation}
    \phi_{\max}\left(z\right)=\pi.
\end{equation}
Outside this interval, the overlap boundary varies linearly with axial coordinate. The corresponding transverse boundary is obtained through linear interpolation between projected corner points:

\begin{equation}
    \label{eq:x_boundary}
    x_{\mathrm{boundary}}\left(z\right)=
    \begin{cases}
        -R
        +
        \left(z-z_{p'_2}\right)
        \dfrac{2R}{z_{p'_3}-z_{p'_2}},
        &
        z<z_{p'_3}
        \\[12pt]
        R,
        &
        z_{p'_3}\le z\le z_{p'_4}
        \\[12pt]
        -R
        +
        \left(z_{p'_1}-z\right)
        \dfrac{2R}{z_{p'_1}-z_{p'_4}},
        &
        z>z_{p'_4}.
    \end{cases}
\end{equation}

The interpolation is exact, because the image of the rim $z=\pm L$ under the central
projection lies in a plane, so the boundary of $C\cap C'$ is linear in $x$
(see \cref{appendix:PG_deriviation}).

The corresponding angular integration limit is then obtained from cylindrical geometry:

\begin{equation}
    \phi_{\max}\left(z\right)
    =
    \arccos\left(
    -\frac{x_{\mathrm{boundary}}\left(z\right)}{R}
    \right)
\end{equation}

Finally, the geometric efficiency is obtained by integrating the solid-angle element over the admissible part of the detector surface. The solid angle subtended at the source $S$ by the region $C\cap C'$ is
\begin{equation}
\Omega(C\cap C')
= \iint_{C\cap C'} \frac{\hat n\cdot\hat r}{r^{2}}\,dA
= \iint_{C\cap C'} \frac{\hat n\cdot\vec r}{r^{3}}\,dA,
\qquad
\vec r = p - S,\quad r = |\vec r|,
\label{eq:solid_angle_def}
\end{equation}
where $\hat n$ is the outward unit normal and $dA$ the area element of the lateral surface. For the parametrisation of \cref{eq:p_param}, the tangent vectors are $\partial_\phi p = (-R\sin\phi,\, R\cos\phi,\, 0)$ and $\partial_z p = (0,\, 0,\, 1)$. Their cross product, $\partial_\phi p\times\partial_z p = R\,(\cos\phi,\, \sin\phi,\, 0)$, gives both the normal and the area element:
\begin{equation}
\hat n = (\cos\phi,\, \sin\phi,\, 0),
\qquad
dA = R\,d\phi\,dz .
\end{equation}
With $\phi_S = 0$, \cref{eq:S_cart} gives the source in Cartesian coordinates as $S = (\rho_S,\, 0,\, z_S)$, so that $\vec r = (R\cos\phi - \rho_S,\; R\sin\phi,\; z - z_S)$ and
\begin{equation}
\hat n\cdot\vec r = R - \rho_S\cos\phi,
\qquad
r^2 = D^2 + (z - z_S)^2,
\end{equation}
where
\begin{equation}
D^2 = R^2 + \rho_S^2 - 2R\rho_S\cos\phi
\label{eq:D2}
\end{equation}
is the squared transverse distance between the source and the surface point. Since $\rho_S < R$, $\hat n\cdot\vec r > 0$ everywhere on $C$ and a ray from an interior source meets the lateral surface at most once, always from the inside, so no contributions cancel. Combining these results with \cref{eq:g_overlap}, the geometric efficiency becomes
\begin{equation}
g=\frac{1}{4\pi}\iint_{C\cap C'} w(\phi,z)\,d\phi\,dz,
\label{eq:2D}
\end{equation}
with the solid-angle integrand
\begin{equation}
w(\phi,z) =
\frac{R\,(R - \rho_S\cos\phi)}
{\left(D^2 + (z-z_S)^2\right)^{3/2}} .
\label{eq:w}
\end{equation}

\subsection{One-dimensional form}

With $u = z - z_S$ denoting the axial offset of the surface point from the source, at fixed $\phi$
\begin{equation}
\int\frac{du}{(D^2+u^2)^{3/2}}=\frac{u}{D^2\sqrt{D^2+u^2}} ,
\end{equation}
so \cref{eq:2D} becomes
\begin{equation}
g=\frac{1}{4\pi}\int_0^{2\pi}\frac{R\,(R-\rho_S\cos\phi)}{D^2}
\left[\frac{u}{\sqrt{D^2+u^2}}\right]_{u_-(\phi)}^{u_+(\phi)}d\phi ,
\label{eq:1Dphi}
\end{equation}
where $u_\pm(\phi)$ are the admissible axial limits, \cref{eq:pg_ulim}.
\Cref{eq:1Dphi} is already a valid one-dimensional form.

It can be simplified further by changing the integration variable from the surface azimuth $\phi$ to
the azimuth $\psi$ of the surface point as seen from the source, for which the prefactor of \cref{eq:1Dphi} is precisely the Jacobian $d\psi/d\phi$.
Denoting by \begin{equation}\label{eq:Dpm_main}
D_\pm(\psi) = \mp\rho_S\cos\psi+\sqrt{R^2-\rho_S^2\sin^2\psi}
\end{equation}
the transverse distances from the source to the wall in the directions $\psi$ and
$\psi+\pi$, the acceptance limits reduce to
\begin{equation}\label{eq:T_main}
T(\psi)=\min\!\left(\frac{a}{D_+},\frac{b}{D_-}\right),
\end{equation}
and the efficiency becomes
\begin{equation}\label{eq:g_T_main}
g=\frac{1}{2\pi}\int_0^{2\pi} f\big(T(\psi)\big)\,d\psi ,
\qquad f(t)=\frac{t}{\sqrt{1+t^2}} .
\end{equation}
The minimum in \cref{eq:T_main} expresses the two-photon condition directly: a
line is accepted when its upward photon rises by less than $a$ and its downward
photon falls by less than $b$. Its two arguments are monotonic in opposite senses
on $(0,\pi)$ and cross at most once, at
\begin{equation}\label{eq:psistar_main}
\psi^*=\arccos\big(\min\{1,\kappa\}\big) ,
\qquad
\kappa = \frac{z_S\sqrt{R^2-\rho_S^2}}{\rho_S\sqrt{L^2-z_S^2}} ,
\end{equation}
with $\psi^*=0$ on the axis. \Cref{eq:g_T_main} can then be written without the minimum as
\begin{equation}\label{eq:g_final_main}
g(\rho_S,z_S)=\frac{1}{\pi}\left[
\int_0^{\psi^*}\frac{b\,d\psi}{\sqrt{b^2+D_-^2(\psi)}}
+\int_{\psi^*}^{\pi}\frac{a\,d\psi}{\sqrt{a^2+D_+^2(\psi)}}\right] .
\end{equation}
The derivation is given in \cref{appendix:PG_deriviation}. Both integrands are
smooth, so \cref{eq:g_final_main} is evaluated by adaptive quadrature at a cost
of a single one-dimensional integration.

The breakpoint $\psi^*$ has a simple geometric meaning. One has $\psi^*=0$, and
hence only the second integral of \cref{eq:g_final_main} contributes, if and only
if
\begin{equation}\label{eq:cone_main}
\rho_S \le \rho_c \equiv \frac{z_S R}{L} ,
\end{equation}
that is, when the source lies inside the cone spanned by the end-cap rim and the
scanner centre. Within this cone the acceptance is limited by the near end alone
and the problem reduces to a single-end expression. Outside it, both ends
contribute, and no single-end formula can be exact. \Cref{eq:cone_main} therefore
delimits the region in which the approximate models of
\cref{subsection:taylor_approx,subsection:cos_approx} can be expected to hold, and it is used
in this sense throughout \cref{section:results}.

\subsection{Model by Lin et al.}\label{subsection:lin_et_al_model}

For comparison with an independently developed analytical approach, the geometric efficiency was also calculated using the model proposed in~\cite{Lin_2026}. This model extends the analytical treatment of \ac{PET} sensitivity developed by~\cite{Bonifacio2023} by considering source displacement in both the axial and transaxial dimensions.

Unlike \ac{PET-GEOM}, which directly evaluates the solid angle of accepted photon-pair directions, the model by \citet{Lin_2026} uses the complementary formulation introduced in \cref{subsection:single_vs_two_geom_eff}. It determines the fraction of emission directions for which at least one photon fails to intersect the lateral surface, corresponding to $(\Omega_0+\Omega_1)/(4\pi)$.

Assume, without loss of generality, that end cap $D_1$ is closer to the source than $D_2$. When viewed from the source, the far end cap $D_2$ projects onto the plane of $D_1$ as a smaller disc $D_2'$. Within a hemisphere facing $D_1$, a photon pair is rejected if the photon travelling towards $D_1$ escapes through that end cap, or if the oppositely directed photon escapes through $D_2$. The corresponding rejected directions are therefore represented by the union $D_1\cup D_2'$.

Because opposite emission directions describe the same photon-pair line with reversed photon assignments, the rejected solid angle in the opposite hemisphere is equal to that in the hemisphere facing $D_1$. Consequently, the complementary expression for the geometric efficiency becomes
\begin{equation}\label{eq:eff_end_caps}
    g = 1 - \frac{\Omega(D_1\cup D_2')}{2\pi},
\end{equation}
where the factor of two accounts for the contributions from both hemispheres.

To compute this solid angle, \citet{Lin_2026} convert the integral over the area of $D_1\cup D_2'$ into an integral along its boundaries. If $D_2'$ lies entirely inside $D_1$ (which corresponds to the inner-cone condition), the boundary is simply the rim of $D_1$. Otherwise, it consists of the arc of $D_1$ and the protruding arc of $D_2'$. In both cases, $g$ reduces to a one-dimensional integral that must be solved numerically (see \cref{appendix:Lin}).

The model is exact and is used here as an independent cross-check of \ac{PET-GEOM}. We implemented it following the description in the original article\footnote{The expression for $\bar{R}_2$ printed in~\cite{Lin_2026} contains a notational inconsistency. Their published results are consistent with the corrected form (see \cref{appendix:Lin}).}.

\subsection{Taylor approximation}\label{subsection:taylor_approx}
To obtain the closed analytical formula as a reference method, we expanded the efficiency in a Taylor series in powers of $\rho_S$.
The details of the derivation are given in \cref{appendix:Taylor}.
Up to fourth order, the series is:
\begin{equation}
g(\rho_S, z_S) = \frac{a}{\sqrt{R^2+a^2}}
+ \frac{3R^2 a}{4\,(R^2+a^2)^{5/2}}\,\rho_S^2
- \frac{15R^2 a\,(4a^2 - 3R^2)}{64\,(R^2+a^2)^{9/2}}\,\rho_S^4
+ \mathcal{O}(\rho_S^6),
\end{equation}
valid for $\rho_S \le \rho_c$ (\cref{eq:cone_main}).

Odd powers vanish after integration over the azimuthal angle $\psi$.
The expansion is exact on the axis ($g(0, z_S)$), and its $\rho_S^2$ coefficient is positive, so for $z_S\neq 0$ the efficiency initially increases with the transaxial offset. In the central plane $\rho_c=0$, the expansion does not apply off the axis, and the efficiency decreases with $\rho_S$ (\cref{tab:efficiency_comparison}).

\subsection{Cosine approximation}\label{subsection:cos_approx}
The fourth model is a two-point angular approximation, used here as a representative of the simple estimators. %

    \begin{equation}
        g(\rho_S,z_S)=
        \begin{cases}
            \frac{L-z_S}{2} \left(\frac{1}{\sqrt{(L-z_S)^2 + (R -\rho_S)^2}} +
            \frac{1}{\sqrt{ (L-z_S)^2 + (R +\rho_S)^2}}\right), & \text{if}\ z_S > \rho_S \frac{L}{R}  \\[2ex]
            \frac{L+z_S}{2} \left(\frac{1}{\sqrt{(L+z_S)^2 + (R +\rho_S)^2}} +
            \frac{1}{\sqrt{ (L+z_S)^2 + (R -\rho_S)^2}}\right), & \text{if}\ z_S \le -\rho_S \frac{L}{R} \\[2ex]
            \frac{1}{2}\left(\frac{L+z_S}{\sqrt{(L+z_S)^2 + (R +\rho_S)^2}} +
            \frac{L-z_S}{\sqrt{ (L-z_S)^2 + (R +\rho_S)^2}}\right), & \text{otherwise}
        \end{cases}
    \end{equation}

The model is exact on the detector axis, where the integrand is independent of $\phi$ and reduces to $\frac{L}{\sqrt{L^2+R^2}}$ at the detector centre. Away from the axis, its accuracy becomes limited.

\section{Implementation}\label{section:implementation}
\subsection{\ac{PET-GEOM} implementation details}\label{subsection:software_details}

The results presented in this work were generated using version 1.0.0 of the library with Python 3.13, NumPy 2.5.2 and SciPy 1.18.1. Unless otherwise stated, calculations use double-precision floating-point arithmetic.

The source code, together with the exact dependency versions is publicly available in the project repository~\citep{PETGeoM}.

\subsection{Numerical implementation}
\label{subsection:numerical_impl}
The implementation follows directly from the geometric construction described in the previous \cref{subsection:pet_geom}. Rather than solving the overlap region analytically, the boundary of the valid integration area is reconstructed from the projections of the four corners of a rectangular $x$--$z$ cross-section.

The integral is evaluated using adaptive two-dimensional numerical Gauss-Kronrod quadrature \texttt{dblquad} from the SciPy library. Both the absolute and relative tolerances were set to their default values of $1.49\times{10}^{-8}$. To assess the sensitivity of the calculated geometric efficiency to the integration tolerances, the calculations were repeated for representative source positions using progressively tighter tolerances, down to $10^{-12}$. Across all tested positions and tolerances, the maximum change in the calculated efficiency relative to the default tolerance was $2.25\times10^{-9}$. Further tightening of the tolerances yielded no substantial change in the calculated efficiencies, indicating that the default integration settings provide sufficient numerical accuracy. Additionally, for a source located at the centre of a detector with $L = 10\ cm$ and $R=20\ cm$, the numerical result agrees with the corresponding analytical solution:
\begin{equation*}
    g(0, 0) = \frac{L}{\sqrt{L^2 + R^2}} = \frac{1}{\sqrt{5}},
\end{equation*}
to better than $10^{-10}$.

The numerical accuracy of the one-dimensional integration method was assessed analogously by varying the absolute and relative error tolerances of the adaptive quadrature algorithm from $1.49\times10^{-8}$ to $10^{-12}$. The estimated absolute integration error was obtained by summing the error estimates returned by \texttt{quad} for the two integrals and dividing their sum by the normalisation factor $\pi$.

For six representative source positions, the calculated geometric efficiencies remained unchanged to the reported numerical precision, with the maximum absolute difference from the result obtained at a tolerance of $10^{-12}$ being $1.33\times10^{-15}$. The estimated absolute integration errors at the default tolerance did not exceed $3.28\times10^{-11}$ and decreased to approximately $10^{-14}$ or below for all tested positions when stricter tolerances were applied. These estimates are several orders of magnitude smaller than the statistical uncertainty of a Monte Carlo estimate based on $10^6$ generated events, indicating that numerical integration errors are negligible for the intended comparison.

\section{Monte Carlo simulations}\label{section:mc_simulations}
\subsection{Simplified MC simulations}\label{subsection:simple_mc}
To provide an independent numerical estimate of the geometric efficiency, a simplified Monte Carlo (MC) simulation was implemented. This simplified model considered only the geometric propagation of the two annihilation photons and did not include photon interactions, energy deposition, detector response, or electronic effects. The geometric efficiency was therefore defined as the fraction of isotropically emitted back-to-back photon pairs for which both photons intersected the active detector surface. For implementation details, see \cref{appendix:simpleMC}.
The geometric efficiency was estimated as
\begin{equation}\label{eqn:mc_efficiency}
    g_{\mathrm{MC}}
    =
    \frac{N_{\mathrm{det}}}{N_{\mathrm{gen}}},
\end{equation}
where $N_{\mathrm{gen}}$ is the total number of generated photon pairs and $N_{\mathrm{det}}$ is the number of pairs for which both photons intersected the finite cylindrical detector surface. Since the simulation contains no stochastic interaction physics, the resulting quantity represents a purely geometric acceptance and is directly comparable with analytical expressions for the geometric efficiency.

The simplified \ac{MC} simulation was used for two types of tests. Firstly, the geometric efficiency of the cylindrical detector for point sources located at arbitrary positions throughout its volume was investigated. For this test, the source was fixed at each considered $(\rho_S,z_S)$ position, and $N_{\mathrm{gen}}=10^6$ isotropically emitted back-to-back photon pairs were generated independently for each source position. The geometric efficiency was then calculated separately for each position according to \cref{eqn:mc_efficiency}. Two tests were performed in this manner, both using a cylindrical detector with a radius of $R = 20$~cm and an \ac{AFOV} of $2L = 20$~cm as the basis. First, three distinct axial offsets $z_S$ of the source were considered, namely $0$~cm, $3$~cm and $6$~cm. For a given $z_S$, the source was then positioned at successive transaxial offsets $\rho_S$, ranging from $0$~cm to $14$~cm in $2$~cm increments. Second, the accuracy of the \ac{PET-GEOM} library was evaluated over a wider range of source positions. In this test, transaxial offsets ranged from $0$~cm to $15$~cm in $1$~cm increments, and the axial offset ranged from $0$~cm to $9$~cm, also in $1$~cm increments.

Next, a test was performed to reproduce the line-source configuration of the analytical model of~\citet{Lin_2026}, in which the point-source efficiency is averaged over a thin rod spanning the axial extent of the detector, and to compare the obtained results with those reported in their study. Note that their simulations and measurements used a 700~mm line source, which extends beyond the 306~mm \ac{AFOV} of their scanner. That configuration was not reproduced here. For the thin-rod source test, the source position was sampled uniformly along the length of the rod. Thus, for each of the $N_{\mathrm{gen}}=10^6$ simulated photon pairs, a source position was randomly selected along the axial extent of the rod, while the radial source position was fixed. An isotropic back-to-back photon pair was then generated from the sampled source position. The resulting efficiency therefore represents the average geometric efficiency over the entire length of the source rod. The efficiency was calculated separately for each of the radial offsets of the rod source from the scanner axis, with the radial displacement varied from $0\%$ to $90\%$ of the detector radius in $5\%$ increments.

In both cases, the geometric efficiency was obtained from the fraction of simulated photon pairs for which both photons intersected the active detector surface. The statistical uncertainty was calculated assuming binomial statistics,

\begin{equation}
    \sigma_{g_{\mathrm{MC}}}
    =
    \sqrt{
    \frac{g_{\mathrm{MC}}\left(1-g_{\mathrm{MC}}\right)}
    {N_{\mathrm{gen}}}
    }.
\end{equation}

For $10^6$ simulated events, the maximum possible absolute statistical uncertainty occurs for $g_{\mathrm{MC}}=0.5$ and is $\sigma_{g_{\mathrm{MC}}}=5\times10^{-4}$, corresponding to an absolute precision of approximately $0.05$ percentage points. For efficiencies smaller than 0.5, the statistical uncertainty is correspondingly lower.
This uncertainty is substantially larger than the maximum variation of $2.25\times10^{-9}$ observed when tightening the numerical integration tolerances, by approximately five orders of magnitude. Thus, the numerical integration error is negligible compared with the statistical uncertainty of the Monte Carlo reference calculations, while the use of $10^6$ events per source position provides sufficient statistical precision for their intended comparison with the numerical geometric efficiency calculations.

\subsection{\ac{GATE}-based MC simulations}\label{subsection:gate_mc}

To check that the position dependence predicted by the geometric models persists when photon transport and detector response are included, \ac{GATE}-based \ac{MC} simulations were performed. They are not used as an accuracy reference, because they also contain non-geometric effects. Simulations described below were performed using \ac{GATE} version 9.4.1 based on Geant4 version 11.2.0.

Two cylindrical detector geometries were implemented for different purposes. The first geometry was used to evaluate the geometric efficiency for a point source positioned at different radial and axial offsets with respect to the detector centre. The second geometry was configured to reproduce the detector geometry considered in~\citet{Lin_2026} and was used to facilitate a direct comparison with their results. Both geometries were constructed as cylindrical shells divided into 24 identical azimuthal LYSO sectors, providing full $360^\circ$ coverage.
A radial detector thickness of 5.0 cm was used to minimise the probability of 511 keV photons traversing the detector without interaction, thereby ensuring that the intrinsic efficiencies of the modelled detectors approached 100\%. The dimensions of both geometries are summarised in \cref{tab:geometries_def}.

\begin{table}
\centering
\caption{Relevant dimensions of both detector geometries used in the tests.}
\label{tab:geometries_def}
     \begin{tabular}{l c c c}
        \toprule
        Simulation setup & Inner radius ($R$) & Length ($2L$) & Radial thickness \\
        \midrule
        Point source & 20.0 cm  & 20.0 cm & 5.0 cm  \\
        Thin rod     & 40.38 cm & 30.6 cm & 5.0 cm  \\
        \bottomrule
    \end{tabular}
\end{table}

For the first simulation setup, a point source emitting two back-to-back photons was placed at positions defined by the radial and axial offsets $(\rho_S,z_S)$. For the second setup, a 0.01 mm line source was used instead, with its length equal to the full length of the detector. Its position was varied along the radial direction to match the configuration of the analytical model of~\citet{Lin_2026}. In both cases, back-to-back decays were simulated using the recently implemented positronium source model described in~\citet{Krzemien_2026}, with the source activity set to $10^5$ Bq. Each simulation was run for 10 s, corresponding to a total of $10^6$ simulated decays.

Electromagnetic interactions were simulated using the \texttt{emlivermore\_polar} physics list available in \ac{GATE}, which includes the polarisation dependence of electromagnetic interactions. Detector signals were processed using the \ac{GATE} digitiser. Energy deposits were first combined within individual detector elements using the adder module, followed by crystal readout. The resulting singles data was used to identify true coincidences based on the event ID associated with each detected photon. Two detected photons were considered to form a true coincidence if they originated from the same primary annihilation event. No energy window or energy-resolution smearing was applied, as the simulations were intended to evaluate the geometric acceptance independently of energy-selection effects.

Finally, the geometric efficiency was estimated as the ratio of detected true coincidence events to the total number of generated primary annihilation events, similarly to \cref{eqn:mc_efficiency}:
\begin{equation}
    g_{GATE} = \frac{N_{\mathrm{det}}}{N_{\mathrm{gen}}}.
\end{equation}

\section{Results}\label{section:results}

\subsection{Validation of \ac{PET-GEOM}}
\label{subsection:validation}
To show the validity of the proposed method for calculating geometric efficiency for a two-gamma, back-to-back emission, the estimations obtained using the \ac{PET-GEOM} library~\citep{PETGeoM} were compared against the simplified \ac{MC} simulation described in \cref{section:mc_simulations}, which served as the reference.
For a given source position, $10^6$ isotropic back-to-back emissions were simulated, and the efficiency was obtained by counting how many of them both photons intersected the detector surface.

The geometric efficiency was calculated for transaxial offsets $\rho_S$ from 0 to 15~cm and axial offset $z_S$ from 0 to 9~cm, both in 1~cm increments, covering both the inner cones and the region outside them.

\begin{figure}[htbp]
    \centering
    \includegraphics[width=.9\textwidth]{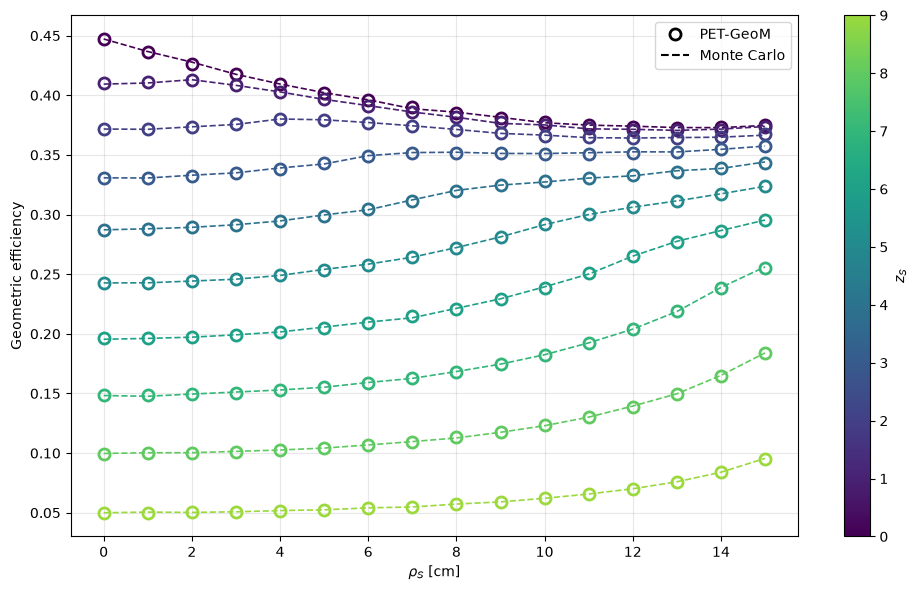}
    \caption{Comparison of geometric efficiency between PET-GeoM and simplified Monte Carlo, obtained for a wide range of source positions.}
    \label{fig:PET-GEM_vs_MC_plots}
\end{figure}

Results are shown in \cref{fig:PET-GEM_vs_MC_plots}.  \ac{PET-GEOM} reproduces the \ac{MC} estimates over the whole range of source positions, including the qualitatively different behaviour at different axial offsets discussed in \cref{subsection:line_sources}.

\begin{figure}[htbp]
 \centering
 \includegraphics[width=.9\textwidth]{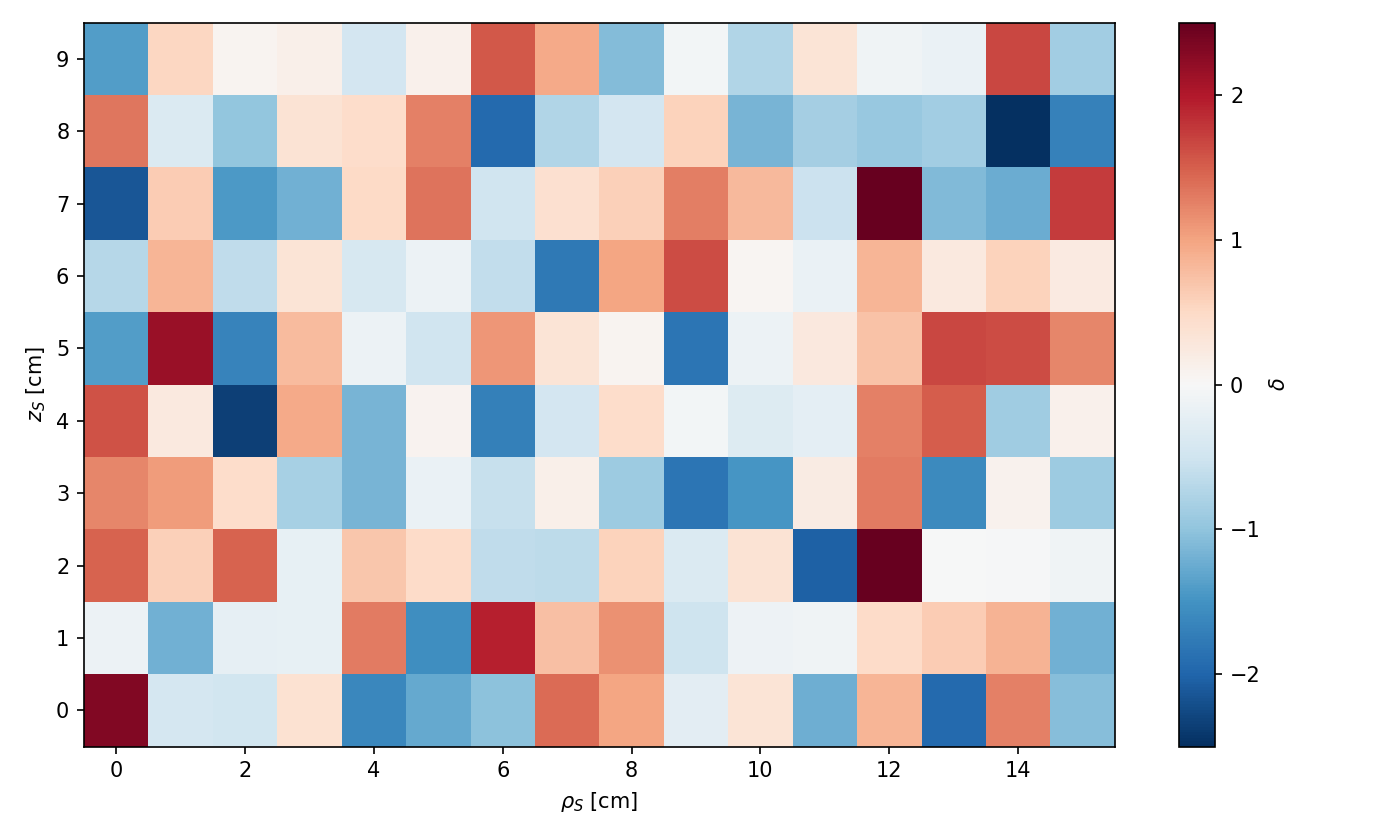}
 \caption{Heatmap of the spatial distribution of the pull values between the PET-GeoM and Monte Carlo estimates of the geometric efficiency for a range of transaxial and axial source offsets.}
 \label{fig:PET-GEM_vs_MC_error_heatmap}
\end{figure}

To quantify the agreement,  for each source position, the difference between the two methods was normalised by the reference Monte Carlo statistical uncertainty:
\begin{equation}
    \delta = \frac{g_{PG} - g_{MC}}{\sigma_{g_{MC}}}
\end{equation}
The heatmap of the spatial distribution of the pull values obtained this way is shown in \cref{fig:PET-GEM_vs_MC_error_heatmap}.
The values show no systematic dependence on the source position, neither inside nor outside the inner cones. Across the 160 investigated source positions, the pull distribution has a mean of $-0.020$ and a standard deviation of $1.094$. Furthermore, $38.13\%$ and $5.00\%$ of the values lie outside $1\sigma$ and $2\sigma$, respectively, compared with the expected fractions of $31.73\%$ and $4.55\%$ for a standard normal distribution. These results indicate that the differences between \ac{PET-GEOM} and the \ac{MC} reference are consistent with the statistical uncertainty of the \ac{MC} calculation, with no evidence of a systematic dependence on the source position. The agreement between \ac{PET-GEOM} and the \ac{MC} reference is limited only by the statistical precision of the latter.

As an independent analytical cross-check, the model of~\citet{Lin_2026}, implemented as described in \cref{subsection:lin_et_al_model}, was also used to calculate the efficiency. For all source positions, the two models agree to within $4.75\times10^{-9}$. This is below the default quadrature tolerance of $1.49\times10^{-8}$ (\cref{subsection:numerical_impl}), so the residual difference reflects the numerical integration of the two models rather than the models themselves.

\subsection{Accuracy of approximate methods}
\label{subsection:approximative}
\label{subsection:results_point_source}
The accuracy of the Taylor and cosine approximations was compared
with the estimations obtained using the \ac{PET-GEOM} together with the GATE-based and simplified \ac{MC} simulations as references.
\ac{GATE} results (\cref{subsection:gate_mc}) are included to show that a full detector simulation reproduces the position dependence. They also contain non-geometric effects and are not used as an accuracy reference.

The point sources were placed at axial offsets $z_S=0$, 3 and 6~cm and transaxial offsets $\rho_S/R$ from 0 to 0.7. For these axial offsets, the boundary of the inner cones lies at $\rho_S/R = 0$, 0.3 and 0.6, respectively.
Results are displayed in~\cref{tab:efficiency_comparison}.

\begin{figure}[htbp]
 \centering
 \includegraphics[width=0.95\textwidth]{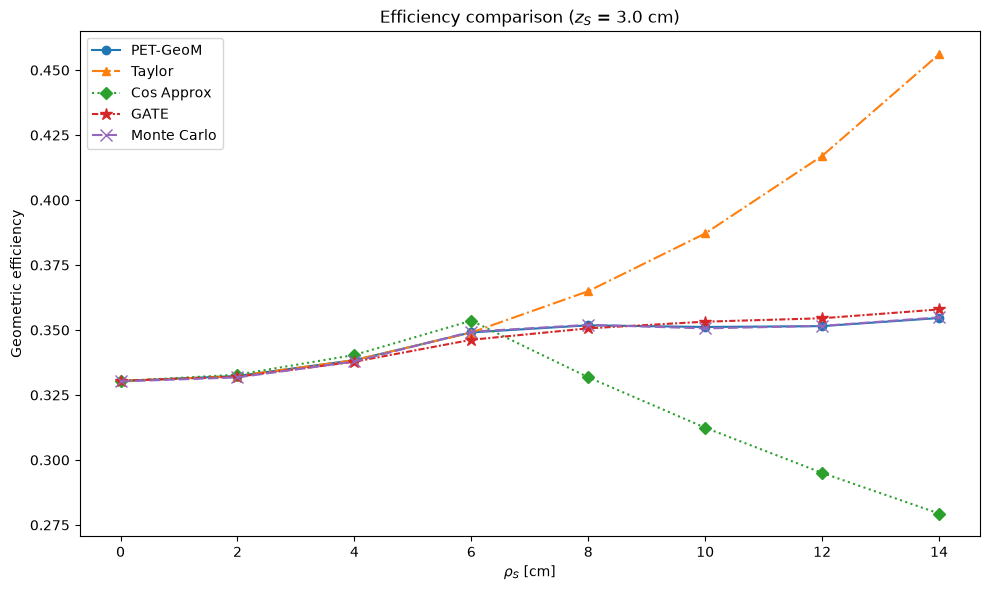}
 \caption{Geometric efficiency as a function of $\rho_S$ for $z_S = 3$~cm, from the data in \cref{tab:efficiency_comparison}.}
 \label{fig:efficiency_graph}
\end{figure}

\begin{table}
    \centering
    \caption{Comparison of geometric efficiency estimates obtained with various methods, with a Monte Carlo estimation as a reference. The point source was positioned at axial offsets $z_S = 0$, $3$, and $6\ cm$, with transaxial offset given as $\rho_S$. Geometric efficiencies are expressed as fractions. Values denoted in blue correspond to positions outside the inner cones ($\rho_S > z_S R/L$).}
    \label{tab:efficiency_comparison}
    \begin{tabular}{l c c c c c}
        \toprule
        $\rho_S/R$ & \ac{PET-GEOM} & Taylor & Cosine & GATE & Monte Carlo \\
        \midrule
        \addlinespace[4pt]
        \multicolumn{6}{c}{$\mathbf{z_S = 0\ cm}$} \\
        \midrule
        0.0                   & 0.4472 & 0.4472 & 0.4472 & 0.4465 & 0.4471 \\
        \blu{0.1} & \blu{0.4265} & \blu{0.4494} & \blu{0.4138} & \blu{0.4363} & \blu{0.4257} \\
        \blu{0.2} & \blu{0.4095} & \blu{0.4559} & \blu{0.3846} & \blu{0.4221} & \blu{0.4101} \\
        \blu{0.3} & \blu{0.3958} & \blu{0.4672} & \blu{0.3590} & \blu{0.4094} & \blu{0.3960} \\
        \blu{0.4} & \blu{0.3852} & \blu{0.4838} & \blu{0.3363} & \blu{0.3981} & \blu{0.3856} \\
        \blu{0.5} & \blu{0.3777} & \blu{0.5062} & \blu{0.3162} & \blu{0.3928} & \blu{0.3778} \\
        \blu{0.6} & \blu{0.3734} & \blu{0.5356} & \blu{0.2983} & \blu{0.3872} & \blu{0.3738} \\
        \blu{0.7} & \blu{0.3732} & \blu{0.5730} & \blu{0.2822} & \blu{0.3880} & \blu{0.3730} \\
        \midrule
        \addlinespace[4pt]
        \multicolumn{6}{c}{$\mathbf{z_S = 3\ cm}$} \\
        \midrule
        0.0                   & 0.3304 & 0.3304 & 0.3304 & 0.3300 & 0.3305 \\
        0.1                   & 0.3323 & 0.3323 & 0.3328 & 0.3315 & 0.3327 \\
        0.2                   & 0.3384 & 0.3384 & 0.3404 & 0.3373 & 0.3377 \\
        0.3                   & 0.3491 & 0.3490 & 0.3536 & 0.3457 & 0.3494 \\
        \blu{0.4} & \blu{0.3518} & \blu{0.3649} & \blu{0.3318} & \blu{0.3501} & \blu{0.3513} \\
        \blu{0.5} & \blu{0.3512} & \blu{0.3872} & \blu{0.3124} & \blu{0.3526} & \blu{0.3512} \\
        \blu{0.6} & \blu{0.3515} & \blu{0.4170} & \blu{0.2950} & \blu{0.3540} & \blu{0.3510} \\
        \blu{0.7} & \blu{0.3546} & \blu{0.4561} & \blu{0.2794} & \blu{0.3574} & \blu{0.3542} \\
        \midrule
        \addlinespace[4pt]
        \multicolumn{6}{c}{$\mathbf{z_S = 6\ cm}$} \\
        \midrule
        0.0                   & 0.1961 & 0.1961 & 0.1961 & 0.1815 & 0.1968 \\
        0.1                   & 0.1975 & 0.1975 & 0.1979 & 0.1830 & 0.1971 \\
        0.2                   & 0.2017 & 0.2017 & 0.2035 & 0.1862 & 0.2019 \\
        0.3                   & 0.2093 & 0.2093 & 0.2134 & 0.1920 & 0.2090 \\
        0.4                   & 0.2211 & 0.2207 & 0.2288 & 0.2015 & 0.2208 \\
        0.5                   & 0.2388 & 0.2371 & 0.2518 & 0.2152 & 0.2386 \\
        0.6                   & 0.2652 & 0.2595 & 0.2856 & 0.2363 & 0.2654 \\
        \blu{0.7} & \blu{0.2863} & \blu{0.2895} & \blu{0.2713} & \blu{0.2583} & \blu{0.2854} \\
        \bottomrule
    \end{tabular}
\end{table}
To better visualise the difference between each method, the data for $z_S = 3$~cm, shown in \cref{tab:efficiency_comparison}, was plotted in \cref{fig:efficiency_graph}.

The results demonstrate that the accuracy of the simplified geometric efficiency models depends on both the transaxial and the axial offset of the source. Inside the inner cones, both approximate methods follow the exact result. The Taylor approximation agrees with \ac{PET-GEOM} with a mean absolute difference of $1.22\times10^{-5}$ for $z_S=3$~cm. The cosine approximation is less accurate and overestimates the efficiency by $4.51\times10^{-3}$ at the cone boundary for $z_S=3$~cm and by $2.05\times10^{-2}$ for $z_S=6$~cm.  Within their applicable region, the accuracy remains high up to a certain radial offset, beyond which it deteriorates as the source approaches the cone boundary. This deterioration is stronger for sources closer to the end caps: at the cone boundary, the absolute error of the Taylor approximation is $4.48\times10^{-5}$ and $5.63\times10^{-3}$ for $z_S=3$ and 6~cm, respectively.

Outside the inner cones, both Taylor and Cosine approximations fail. For $\rho_S>z_S R/L$,
the Taylor approximation increasingly overestimates the efficiency, and the
cosine approximation underestimates it.

\subsection{Point and line sources}
\label{subsection:line_sources}
\begin{figure}[htbp]
     \centering
     \includegraphics[width=.9\textwidth]{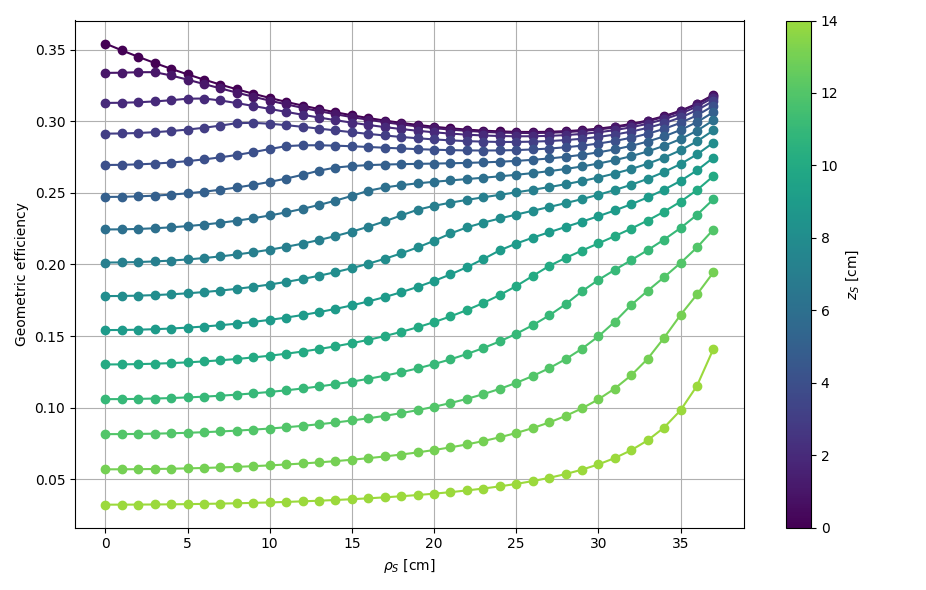}
     \caption{Geometric efficiency as a function of transaxial and axial offsets of a point source. The geometry used here was a cylindrical detector with inner radius $R = 40.38$~cm and \ac{AFOV} $2L = 30.6$~cm. }
     \label{fig:Lin_geom_point_source_eff}
\end{figure}

As shown in \cref{fig:Lin_geom_point_source_eff}, for a point source, the dependence of the geometric efficiency on the transaxial offset changes qualitatively with the axial position. In the central plane, $z_S=0$, the efficiency initially decreases with increasing transaxial offset, reaches a minimum, and subsequently increases as the source approaches the detector surface. In planes close to the detector end caps, the efficiency increases monotonically with transaxial offset, with the rate of increase becoming progressively steeper as the source approaches the detector surface. At intermediate axial positions, the dependence is more complex: the efficiency initially increases slowly, followed by a change in slope and a period of slight decrease or near-constant behaviour. At larger transaxial offsets, the efficiency then begins to increase again, with the rate of increase becoming progressively steeper as the source approaches the detector surface.
For the geometries considered here, the efficiency is highest at the centre of the scanner.

\begin{figure}[htbp]
 \centering
 \includegraphics[width=.9\textwidth]{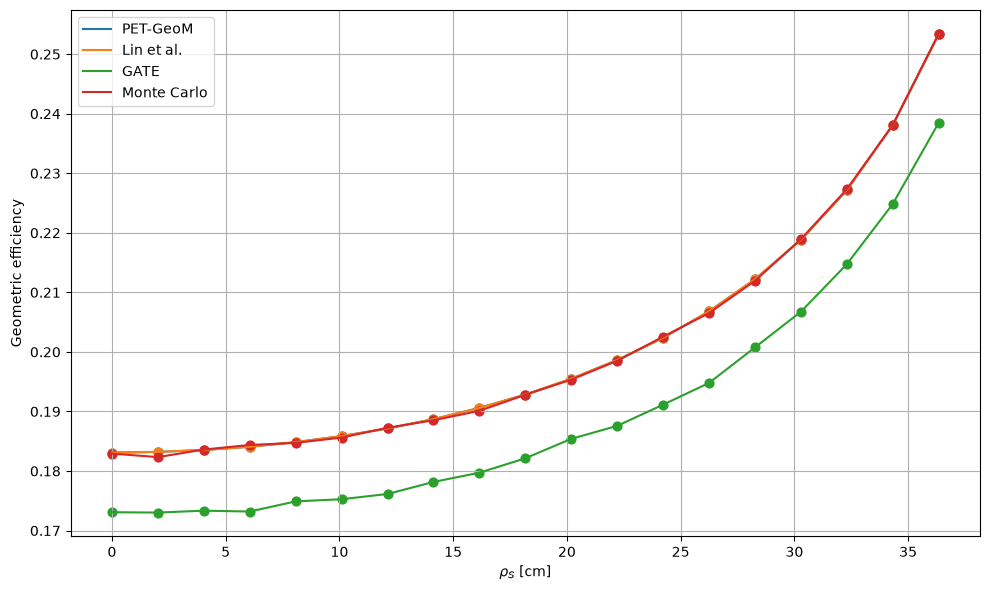}
 \caption{Geometric efficiency as a function of transaxial source offset for a thin source rod extending over the full axial length of the detector, calculated using the model proposed by \citet{Lin_2026}, \ac{PET-GEOM}, the simplified \ac{MC} simulation, and \ac{GATE} simulations.}
 \label{fig:PET-GEM_vs_Lin_rod_source}
\end{figure}

To relate these results to the line-source configuration considered by~\citet{Lin_2026}, the geometric efficiency of a thin rod extending over the full axial length of the detector was calculated as a function of its transaxial offset.
The test reproduces the detector geometry and the line-source configuration of the analytical model of~\citet{Lin_2026}, thereby allowing the proposed model to be assessed under conditions corresponding to a previously investigated detector geometry. Unlike in their simulations and measurements, the rod does not extend beyond the \ac{AFOV}.
The basis for this test was a cylindrical detector with an inner radius $R = 40.38\ cm$ and \ac{AFOV} $2L = 30.6\ cm$. For each rod position, the geometric efficiency was calculated using \ac{PET-GEOM}, by averaging the point-source efficiency over the axial coordinate, and using the model of \citet{Lin_2026},  while an equivalent configuration was simulated using the simplified \ac{MC} and \ac{GATE}. The results of this validation are shown in \cref{fig:PET-GEM_vs_Lin_rod_source}.

\ac{PET-GEOM} and the model of \citet{Lin_2026} coincide and agree with the \ac{MC} reference. The \ac{GATE} values follow the same trend but lie about 0.01 lower, reflecting non-geometric effects. The rod efficiency increases monotonically with the transaxial offset, from $0.183$ on the axis to $0.253$ at $\rho_S=0.9R$.

\subsection{Organ-like regions and transaxial offset}
    \label{subsection:results_average_eff}
The point source and the rod source represent two limiting cases. For the latter configuration, which is described in \cref{subsection:line_sources}, \citet{Lin_2026} reported an increase in sensitivity with transaxial offset.
To determine which of them applies to a patient, the geometric efficiency was averaged over a cylindrical region of radius $r_o$ and axial length $2l_o$, centred in the central plane, with its axis shifted transaxially by $d$,
\begin{equation}\label{eq:g_organ}
    \bar{g}(d) = \frac{1}{V_o}\int_{V_o(d)} g(\rho_S, z_S)\,\mathrm{d}V,
    \qquad
    \Delta(d) = \frac{\bar{g}(d)}{\bar{g}(0)} - 1 .
\end{equation}
The integral was evaluated with \ac{PET-GEOM}. As a cross-check, $\bar{g}(d)$ was also estimated with the simplified \ac{MC} simulation (\cref{subsection:simple_mc}), with source positions sampled uniformly within the region and $10^8$ photon pairs per configuration. For all 365 configurations, the two agreed within the statistical uncertainty of the \ac{MC}, which is below 0.06 percentage points in $\Delta$. Idealised geometries with the dimensions of four reference \ac{PET} systems were used (\cref{tab:geometries_config}). The fraction of the \ac{AFOV} covered by the region, $l_o/L$, was varied from 0.1 to 1. $l_o/L=1$ corresponds to the full-\ac{AFOV} rod of \cref{subsection:line_sources}. As a realistic example, the ascending colon of the adult stylised phantom of the \acf{MIRD} type~\citep{Cristy1980} was used for the clinical scanners. In this phantom, organs are represented by simple geometric bodies. The ascending colon (wall and contents) is a circular cylinder ($r_o=2.5$~cm, length 9.55~cm) parallel to the scanner axis and thus matches the region model directly. Its axis lies about 9~cm from the trunk axis, so $d\approx9$~cm already for a patient centred in the scanner.

\begin{table}
    \centering
    \caption{Idealised cylindrical scanner geometries with the dimensions of the reference \ac{PET} systems.}
    \label{tab:geometries_config}
     \begin{tabular}{l c c c c}
        \toprule
          & Philips Vereos & Siemens Vision & Quadra & uExplorer \\
        \midrule
        Length [cm]         & 16.4   & 26.1    & 106.0  & 194.0 \\
        Inner radius [cm]   & 38.2   & 39.0    & 41.0   & 39.3 \\
        2L/R                & 0.4293 & 0.6692  & 2.5854 & 4.9364 \\
        \bottomrule
    \end{tabular}
\end{table}

\begin{figure}[htbp]
 \centering
 \includegraphics[width=.9\textwidth]{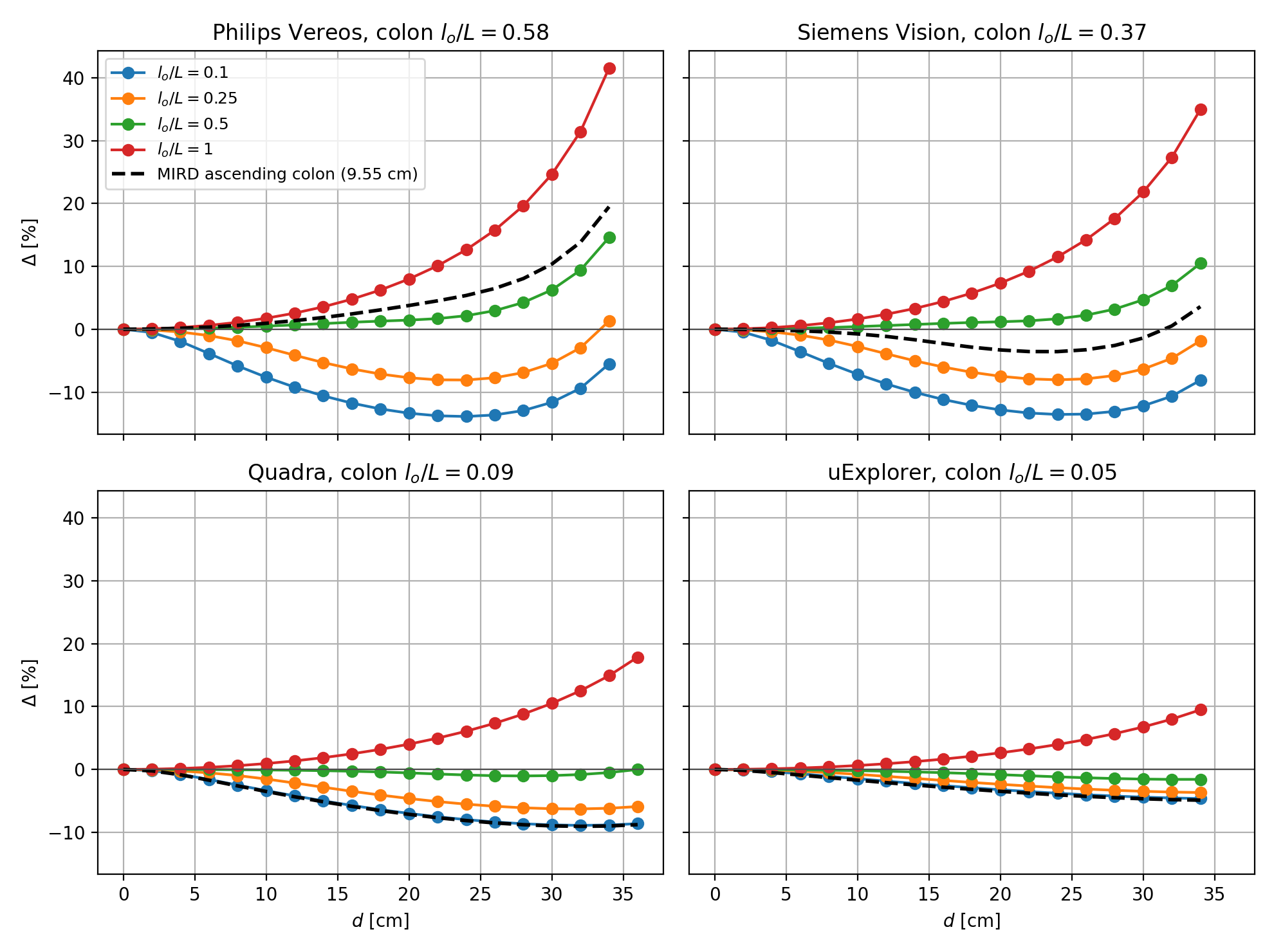}
 \caption{Relative change $\Delta$ of the region-averaged geometric efficiency, \cref{eq:g_organ}, with transaxial offset $d$ for four clinical scanner geometries ($r_o=2.5$~cm). Coloured lines: regions covering a fraction $l_o/L$ of the \ac{AFOV}. dashed line: \acs{MIRD} ascending colon.}
 \label{fig:organ_offset_clinical}
\end{figure}

The sign of $\Delta$ is set by $l_o/L$ (\cref{fig:organ_offset_clinical}). At $d=10$~cm, a region with $l_o/L=0.1$ loses 7.6\% of its efficiency in the Vereos geometry and 3.4\% in the Quadra geometry, whereas a region spanning the full \ac{AFOV} gains 1.8\% and 0.9\%, respectively. The change of sign occurs close to $l_o/L=0.5$ in all four geometries, slightly below it in the short-\ac{AFOV} scanners and slightly above it in the long-\ac{AFOV} ones. Consequently, the same organ behaves differently in different scanners: at $d=10$~cm, the efficiency for the colon increases by 0.95\% in the Vereos ($l_o/L=0.58$) but decreases by 0.7\%, 3.5\% and 1.7\% in the Vision, Quadra and uExplorer geometries ($l_o/L=0.37$, $0.09$ and $0.05$). In long-\ac{AFOV} scanners, where an organ covers only a small fraction of the \ac{AFOV}, a transaxial offset therefore reduces its geometric efficiency.

\subsection{Computational time}
    \label{subsection:results_comp_time}

The computational time required by \ac{PET-GEOM} was compared with that of the simplified Monte Carlo implementation described in \cref{subsection:simple_mc} and of the GATE simulations described in \cref{subsection:gate_mc}.
For each benchmark configuration, the source position was randomly sampled before each measurement within the region defined by
\begin{equation*}
    0 \leq \rho_S < 0.9R,
    \qquad
    -0.9L < z_S < 0.9L.
\end{equation*}
for the detector geometry with $R=20$~cm and $2L=20$~cm.
The same set of source positions was used for all methods within each measurement, ensuring that the comparison was performed over the same set of geometric configurations.
The simplified \ac{MC} simulation was run with $10^5$ and $10^6$ photon pairs per source position. The latter corresponds to the statistics used in \cref{subsection:validation}.
The \ac{GATE} simulations used $10^6$ primary annihilations per position.

Prior to the measurements, 20 warm-up evaluations were performed for both \ac{PET-GEOM} and the simplified \ac{MC} simulation and excluded from the reported execution times. The execution time of each individual calculation was measured using a high-resolution performance counter on a single CPU core. For each configuration, the same procedure was repeated over multiple independent runs, and the mean execution time was reported. The number of such runs, denoted as \textit{Number of Runs}, was selected separately for each configuration. The resulting mean execution times, together with the standard error of the mean (SEM) and median values, are summarised in \cref{tab:average_comp_time}.

\begin{table}
    \centering
    \caption{Execution time per source position for the PET-GeoM library and for the simplified and GATE Monte Carlo estimations: mean, standard error of the mean (SEM) and median over the given number of runs.}
    \label{tab:average_comp_time}
     \begin{tabular}{l c c c c c}
        \toprule
        Model & Sim. Events & Number of Runs & Mean Time [s] & SEM [s] & Median [s] \\
        \midrule
        PET-GeoM (1D)&    —    & $10^3$ & $5.29\times10^{-5}$ & $6.19\times10^{-7}$  & $4.4\times10^{-5}$  \\
        PET-GeoM (2D)&    —    & $10^3$ & $0.727$   & $0.044$  & $0.314$  \\
        Monte Carlo  & $10^5$  & $10^4$ & $2.13$    & $0.017$  & $0.745$  \\
        Monte Carlo  & $10^6$  & $10^3$ & $12.2$    & $0.37$   & $7.31$  \\
        GATE         & $10^6$  & 150    & $373$     & $33.4$   & $210$ \\
        \bottomrule
    \end{tabular}
\end{table}

The one-dimensional formulation, \cref{eq:g_final_main}, is about $4\times10^{4}$ times faster than the simplified \ac{MC} with $10^5$ photon pairs, whose statistical uncertainty reaches $1.6\times10^{-3}$, and about $2\times10^{5}$ times faster than with $10^6$ pairs (\cref{tab:average_comp_time}). Since the cost of the \ac{MC} estimate scales as $\sigma^{-2}$, approaching the numerical precision of \ac{PET-GEOM} (below $10^{-10}$, \cref{subsection:numerical_impl}) by sampling would be impractical. The two-dimensional formulation and \ac{GATE} are listed for reference. The \ac{GATE} time includes full photon transport and detector response, and indicates the cost of obtaining the geometric efficiency from a complete simulation, as is common in practice.

\section{Discussion}
    \label{section:discussion}

\subsection{Model applicability region}
    \label{subsection:discussion_applicability}

The approximate models are reliable only inside the two inner cones, $\rho_S\le |z_S|R/L$ (\cref{eq:cone_main}), which together occupy one-third of the scanner volume. The reason is geometric. Inside a cone, the nearer end cap alone limits the admissible directions, and the efficiency reduces to a single-end expression (\cref{eq:g_final_main} with $\psi^*=0$), which the Taylor approximation expands and the cosine approximation samples. Outside the cones, the far end cap limits the acceptance of lines whose upward photon is emitted towards the nearer wall ($\psi<\psi^*$), and no single-end description can be exact. The validity of these approximations is therefore set by the relative transaxial and axial position of the source, not by the magnitude of the transaxial offset alone.

\subsection{Two-photon acceptance and transaxial position}
    \label{subsection:discussion_transaxial}

The single-photon solid angle and the two-photon efficiency coincide only at the scanner centre. Off the axis, the oppositely directed photon must also reach the detector. In the central plane this makes $g$ decrease over most of the transaxial range (\cref{tab:efficiency_comparison}), a behaviour that cannot be inferred from the single-photon solid angle.

\subsection{Line sources and implications for sensitivity characterisation}
Offsetting a patient transaxially does not generally increase the counts from a given region. This differs from the conclusion of~\citet{Lin_2026}, and their results require particular consideration. With the corrected radius of $D_2'$ the model is exact and agrees with \ac{PET-GEOM} across all considered tests. The difference lies in the interpretation.
The reported observation that sensitivity increases with transaxial offset applies to the specific source configuration considered in their validation and should not be interpreted as a general property of cylindrical \ac{PET} geometry. In their analysis, the source was represented by a line source extending along the detector axis, and the reported sensitivity for a given radial position was therefore effectively averaged over the axial coordinate. The point-source results presented here show that the dependence on transaxial offset changes with the axial position (\cref{fig:Lin_geom_point_source_eff}).  The increase observed for the axially extended source results from averaging the point-source efficiency over the range of axial positions occupied by the rod, in which positions close to the end caps contribute the strongest increase. Whether an offset increases the counts from a given region therefore depends on the fraction of the \ac{AFOV} it covers (\cref{subsection:results_average_eff}). For organ-sized regions in long-\ac{AFOV} scanners, the efficiency decreases. For a point source in the central plane, the reduction reaches 12\% at a 10~cm offset in a scanner with the dimensions of the Philips Vereos.

The same distinction applies to standard sensitivity measurements, which use a line source at the centre and at a 10~cm transaxial offset~\citep{NEMA:2018}. For a rod covering the full \ac{AFOV}, the geometric component alone predicts a small increase at 10~cm, between 0.6\% (uExplorer) and 1.7\% (Vereos) for the scanners considered here. The NEMA source is 70~cm long. In the Quadra and uExplorer geometries it covers only part of the \ac{AFOV}, and the predicted change becomes $+0.4\%$ and $-0.5\%$, respectively. In the Vereos and Vision geometries it extends beyond the \ac{AFOV}, which is not modelled here. Effects not included in the geometric efficiency, such as the dependence of intrinsic efficiency on the incidence angle, gaps between detector modules, and depth of interaction, are of a comparable magnitude and may reverse this trend. This may explain why both higher and lower off-centre sensitivities have been reported for different systems.

\subsection{Limitations}
    \label{subsection:discussion_limitations}

\ac{PET-GEOM} describes only the geometric component of the detection efficiency, which is often factorised as
\begin{equation*}
\eta_{det} = g \cdot\varepsilon\ \cdot f \cdot F,
\end{equation*}
where $\varepsilon$ is the intrinsic efficiency, $f$ the fraction of detector signals accepted by the acquisition system (e.g.\ by the energy window), and $F$ accounts for attenuation and scattering between the source and the detector. In a real \ac{PET} system these factors are not separable: $\varepsilon$ and $f$ depend on the photon incidence angle and therefore on the source position and \ac{LOR} geometry. The model also assumes an ideal cylinder with a continuous active lateral surface and neglects detector segmentation, gaps, finite crystal size, depth of interaction, coincidence timing and electronic effects. Some scanners additionally restrict the accepted \ac{LOR} angles. The geometric efficiency is therefore not a prediction of the measured sensitivity, but an exact reference for the ideal two-photon acceptance and one component of a more complete sensitivity model.

\section{Conclusions}
    \label{section:conclusions}

We presented \ac{PET-GEOM}, an exact method and open-source Python library for calculating the two-photon geometric efficiency of a cylindrical \ac{PET} detector for a point source at an arbitrary position.  The method is based on the central projection of the detector surface through the source and reduces to a one-dimensional integral.

The proposed method was validated against a simplified \ac{MC} reference over a wide range of transaxial and axial source offsets, with agreement within the statistical uncertainty of the \ac{MC} calculation, while the numerical integration error of \ac{PET-GEOM} was negligible in comparison. \ac{GATE} simulations including detector response reproduce the same position dependence. The method also agrees with the independently developed model of~\citet{Lin_2026} for all source positions considered, including the axially extended thin-rod source configuration.

Using the \ac{PET-GEOM} formulation for the two-photon geometric efficiency of a point source, we showed that it jointly depends on the transaxial and axial coordinates of the source. In the central plane, it generally decreases with transaxial offset, whereas close to the end caps it increases, so its behaviour cannot be inferred from the single-photon solid angle. The increase with transaxial offset reported for line sources results from averaging over the axial coordinate and is not a general property of cylindrical \ac{PET} geometry. Whether an offset increases the efficiency for an extended region depends on the fraction of the \ac{AFOV} it covers. For organ-sized regions in long-\ac{AFOV} scanners, it decreases.

The computationally efficient Taylor and cosine approximations are, by construction, applicable only inside the two inner cones defined by the end-cap rims and the scanner centre, which occupy one-third of the scanner volume. Their accuracy is therefore set not only by the transaxial offset but also by the relative transaxial and axial position of the source.

With a cost of about $50~\mu$s per evaluation and no statistical uncertainty, \ac{PET-GEOM} is suited to applications requiring many evaluations, such as sensitivity-map generation and scanner-geometry studies.

The two-dimensional formulation underlying \ac{PET-GEOM} is not specific to the cylinder.
The key observation is that a line through the source is accepted if both of its ends intersect the detector's active surface. Consequently, the two-photon efficiency is given, in general, by the solid angle of the overlap between the active surface and its central projection through the source, divided by $4\pi$. While the one-dimensional reduction relies on the circular cross-section,
this construction can, in principle, be applied to any convex detector surface enclosing the source, including polygonal rings of flat detector modules and detectors with gaps, which can be represented as excluded regions of the active surface.
For polygonal geometries, central projection maps the module boundaries onto straight lines, so the overlap consists of planar polygons whose solid angles are known in closed form.

\ack{The authors acknowledge the support by the Foundation for Polish Science through the FIRST TEAM FENG.02.02-IP.05-0152/23 programme co-financed by the European Union under the European Funds for Smart Economy 2021-2027 (FENG). The work is co-financed by the Polish National Agency for Academic Exchange in the frame of the projects BPN/BFR/2025/1/00036/U/00001 and BPN/BAT/2025/1/00009 and co-supported by the Austrian Ministry for Women, Science and Research (project WTZ PL 11/2026).
This work was partially supported by the “PHC POLONIUM” program (project number: 55194PB), funded by the French Ministry for Europe and Foreign Affairs, the French Ministry for Higher Education, Research and Space, and the Polish NAWA.
This work was completed with resources provided by the Świerk Computing Centre at the National Centre for Nuclear Research.}

\section*{Use of generative AI}
The possibility of reducing the surface integral, \cref{eq:2D}, to a
one-dimensional form, by integrating analytically over the axial coordinate,
was identified by the authors. The closed-form expressions for the boundaries
of the admissible region and the subsequent
derivation of the one-dimensional form were carried out with the assistance
of the generative AI model Claude Opus 5.5 (Anthropic). The authors verified
the derivation analytically and numerically against the two-dimensional
integration and the Monte Carlo reference (\cref{subsection:validation}).

\section*{Data availability}
{The \ac{PET-GEOM} library (version 1.0.0) is openly available at \url{https://github.com/WBZdeb/PET-GeoM}~\citep{PETGeoM}. The Monte Carlo simulation data that support the findings of this study are available upon reasonable request.}

\appendix
\crefalias{section}{appendix}
\section{Reduction of the \ac{PET-GEOM} integral to one dimension}
\label{appendix:PG_deriviation}

This appendix completes the reduction of \cref{eq:1Dphi} to the compact form of \cref{eq:g_final_main}. Throughout, the notation of \cref{section:notation} is used, with $\rho_S<R$ and $0\le z_S<L$ (for $z_S<0$ replace $z_S$ by $|z_S|$). We abbreviate the axial distances from the source to the near and far end planes, and the power of the source with respect to the detector circle, as
\begin{equation}\label{eq:pg_abc}
a = L-z_S, \qquad b = L+z_S, \qquad c^2 = R^2-\rho_S^2 ,
\end{equation}
and write $u = z-z_S$ for the axial offset of a surface point from the source.

\subsection{The acceptance condition in closed form}

Let the line through $S$ and a surface point $p(\phi,z)$ meet the cylinder a second time at $p'$. In the transverse plane $S$ lies on the chord $pp'$, so by the power of a point with respect to the circle of radius $R$,
\begin{equation}
|Sp|_{xy}\,|Sp'|_{xy} = R^2-\rho_S^2 = c^2 .
\end{equation}
Hence $p' = S-k(\phi)\,(p-S)$ with
\begin{equation}\label{eq:pg_k}
k(\phi) = \frac{c^2}{D^2(\phi)} ,
\end{equation}
and, since the axial coordinate varies linearly along the line, $z' = z_S-k u$. The point $p$ belongs to $C\cap C'$ if and only if $|z|\le L$ and $|z'|\le L$, that is
\begin{equation}
-b\le u\le a , \qquad -\frac{a}{k}\le u\le \frac{b}{k} .
\end{equation}
For each $\phi$ the admissible axial range is therefore the single interval $u\in[u_-(\phi),u_+(\phi)]$ with
\begin{equation}\label{eq:pg_ulim}
u_+=\min\!\left(a,\ \frac{b}{k}\right), \qquad
u_-=-\min\!\left(b,\ \frac{a}{k}\right) ,
\end{equation}
which always contains $u=0$, so that no case distinction is required.

\Cref{eq:pg_ulim} also shows that the boundary of $C\cap C'$ is linear in $x$: eliminating $\phi$ from the image of the rim $z=L$ gives the plane
\begin{equation}
z = (2z_S-L)+\frac{2\rho_S (L-z_S)}{R^2-\rho_S^2}\,(x-\rho_S) ,
\end{equation}
and analogously for $z=-L$. The linear interpolation used in \cref{eq:x_boundary} is therefore exact, not an approximation.

\subsection{Change to the azimuth seen from the source}

Let $\psi$ be the azimuth of $p$ as seen from $S$, defined by $\tan\psi = R\sin\phi/(R\cos\phi-\rho_S)$. With $x=R\cos\phi-\rho_S$ and $y=R\sin\phi$,
\begin{equation}\label{eq:pg_jac}
\frac{d\psi}{d\phi} = \frac{x\,y'-y\,x'}{x^2+y^2} = \frac{R\,(R-\rho_S\cos\phi)}{D^2} > 0 ,
\end{equation}
so $\phi\mapsto\psi$ is a monotonic bijection of the circle and the prefactor of \cref{eq:1Dphi} is exactly this Jacobian.

The ray $S+t(\cos\psi,\sin\psi)$ meets the detector circle when $t^2+2\rho_S t\cos\psi-c^2=0$, so the transverse distances from the source to the wall in the directions $\psi$ and $\psi+\pi$ are
\begin{equation}\label{eq:pg_Dpm}
D_\pm(\psi) = \mp\rho_S\cos\psi+\sqrt{R^2-\rho_S^2\sin^2\psi} ,
\qquad D_-(\psi)=D_+(\psi+\pi), \qquad D_+D_-=c^2 .
\end{equation}
By construction $D(\phi)=D_+(\psi)$, whence $1/(kD)=D/c^2=1/D_-(\psi)$. Dividing the limits \cref{eq:pg_ulim} by $D$ gives
\begin{equation}\label{eq:pg_T}
\frac{u_+}{D}=\min\!\left(\frac{a}{D_+},\frac{b}{D_-}\right)\equiv T(\psi) ,
\qquad
\frac{u_-}{D}=-\min\!\left(\frac{b}{D_+},\frac{a}{D_-}\right)=-T(\psi+\pi) .
\end{equation}
Since $u/\sqrt{D^2+u^2}=f(u/D)$ with $f(t)=t/\sqrt{1+t^2}$, \cref{eq:1Dphi} becomes
\begin{equation}\label{eq:pg_Tform}
g=\frac{1}{4\pi}\int_0^{2\pi}\Big[f\big(T(\psi)\big)+f\big(T(\psi+\pi)\big)\Big]\,d\psi
 =\frac{1}{2\pi}\int_0^{2\pi} f\big(T(\psi)\big)\,d\psi ,
\end{equation}
the second equality following from the periodicity of the integrand.

This form has a direct interpretation. Writing $u=D\tan\lambda$ with $\lambda$ the elevation angle, one finds $w\,d\phi\,du=\cos\lambda\,d\lambda\,d\psi$, the solid-angle element in spherical coordinates centred on the source. A line of elevation $\lambda>0$ is accepted when its upward photon rises by less than $a$ and its downward photon falls by less than $b$, that is when $\tan\lambda\le T(\psi)$. $g$ is thus the accepted fraction of the upper hemisphere of directions.

\subsection{Breakpoint and final form}

Because $D_\pm$ depend on $\cos\psi$ and $\sin^2\psi$ only, the integrand of \cref{eq:pg_Tform} is even in $\psi$ and the range may be halved. On $(0,\pi)$,
\begin{equation}
\frac{dD_+}{d\psi}=\rho_S\sin\psi\left[1-\frac{\rho_S\cos\psi}{\sqrt{R^2-\rho_S^2\sin^2\psi}}\right]>0
\end{equation}
since $R>\rho_S$, so $D_+$ increases from $R-\rho_S$ to $R+\rho_S$ while $D_-=c^2/D_+$ decreases. The two arguments of the minimum in \cref{eq:pg_T} are therefore monotonic in opposite senses and cross at most once. The crossing $aD_-=bD_+$ gives $D_+^{*2}=c^2a/b$ and, inserted into $D_+^2+2\rho_S D_+\cos\psi-c^2=0$,
\begin{equation}\label{eq:pg_psistar}
\kappa \equiv \frac{c\,(b-a)}{2\rho_S\sqrt{ab}}
            = \frac{z_S\sqrt{R^2-\rho_S^2}}{\rho_S\sqrt{L^2-z_S^2}} ,
\qquad
\psi^* = \arccos\big(\min\{1,\kappa\}\big) ,
\end{equation}
with $\psi^*=0$ for $\rho_S=0$, for which $\kappa$ is singular but the source lies on the axis.
Splitting \cref{eq:pg_Tform} at $\psi^*$ and evaluating $f$ on each branch yields the final result:
\begin{equation}\label{eq:pg_final}
g(\rho_S,z_S)=\frac{1}{\pi}\left[
\int_0^{\psi^*}\frac{b\,d\psi}{\sqrt{b^2+D_-^2(\psi)}}
+\int_{\psi^*}^{\pi}\frac{a\,d\psi}{\sqrt{a^2+D_+^2(\psi)}}\right]
\end{equation}
Both integrands are smooth on their intervals, so adaptive Gauss--Kronrod quadrature converges rapidly.

Evaluating \cref{eq:pg_T} at $\psi=0$ shows that $a/D_+\le b/D_-$, and hence $\psi^*=0$, if and only if $a(R+\rho_S)\le b(R-\rho_S)$, that is
\begin{equation}\label{eq:pg_cone}
\rho_S \le \frac{z_S R}{L} \equiv \rho_c .
\end{equation}
Inside this cone the near end alone limits the acceptance at every azimuth and only the second integral of \cref{eq:pg_final} survives. Outside it the far end limits the acceptance for $\psi<\psi^*$, that is for lines whose upward photon is emitted towards the nearer wall, so that the downward photon crosses the scanner the long way. The single-end expression obtained inside the cone is what the approximations of \cref{subsection:taylor_approx,subsection:cos_approx} expand or sample, which explains why both deteriorate once $\psi^*>0$.

Several limits follow immediately from \cref{eq:pg_final}. On the axis ($\rho_S=0$) one has $\psi^*=0$ and $D_\pm=R$, so that $g=a/\sqrt{a^2+R^2}$, the known result for an axially displaced source, which reduces at the detector centre to $g=L/\sqrt{L^2+R^2}$. For a source in the transaxial plane through the centre ($z_S=0$, $\rho_S>0$) the breakpoint lies at $\psi^*=\pi/2$ and, since $D_-(\psi)=D_+(\pi-\psi)$, the two integrals of \cref{eq:pg_final} are equal.

\section{Taylor expansion derivation}\label{appendix:Taylor}
Inside the cone $\rho_S\le\rho_c$ (\cref{eq:pg_cone}) one has $\psi^*=0$, and only the second integral of \cref{eq:pg_final} contributes:
\begin{equation}
g(\rho_S,z_S)=\frac{1}{\pi}\int_0^{\pi}\frac{a\,d\psi}{\sqrt{a^2+D_+^2(\psi)}} ,
\end{equation}
where, from \cref{eq:pg_Dpm},
\begin{equation}
\begin{split}
D_+^2(\psi) &= R^2 + \rho_S^2\cos 2\psi - 2\rho_S\cos\psi\sqrt{R^2-\rho_S^2\sin^2\psi} \\
&= R^2 - 2R\rho_S\cos\psi + \rho_S^2\cos 2\psi + \frac{\rho_S^3}{R}\sin^2\psi\cos\psi + \mathcal{O}(\rho_S^5) .
\end{split}
\end{equation}
Writing $a^2+D_+^2 = (R^2+a^2)(1+\epsilon)$ with $\epsilon = (D_+^2-R^2)/(R^2+a^2) = \mathcal{O}(\rho_S)$, the integrand is expanded as
\begin{equation}
\frac{a}{\sqrt{R^2+a^2}}\left(1-\frac{\epsilon}{2}+\frac{3\epsilon^2}{8}-\frac{5\epsilon^3}{16}+\frac{35\epsilon^4}{128}+\mathcal{O}(\epsilon^5)\right) .
\end{equation}
Since $D_+(\pi-\psi)$ at $\rho_S$ equals $D_+(\psi)$ at $-\rho_S$, the integral is even in $\rho_S$ and all odd powers vanish. Collecting the terms up to $\rho_S^4$ and integrating over $\psi$ gives
\begin{equation}
g(\rho_S,z_S) = \frac{a}{\sqrt{R^2+a^2}}
+ \frac{3R^2 a}{4\,(R^2+a^2)^{5/2}}\,\rho_S^2
- \frac{15R^2 a\,(4a^2-3R^2)}{64\,(R^2+a^2)^{9/2}}\,\rho_S^4
+ \mathcal{O}(\rho_S^6) .
\end{equation}
The leading term is the exact on-axis result. Outside the cone $\psi^*>0$, the first integral of \cref{eq:pg_final} no longer vanishes, and the expansion does not apply.

\section{Model of Lin et al. in the present notation}
\label{appendix:Lin}

The formulation of~\cite{Lin_2026} provides the geometric efficiency (there called geometric sensitivity) as a function of the source position in the cylindrical detector volume. The estimation requires the numerical calculation of a one-dimensional integral. The model is derived from the solid angle subtended by the detector at the source, with the complete derivation and validation provided in the original publication. Here, only the formulation relevant to its implementation in the present study is described.

The model evaluates \cref{eq:eff_end_caps} directly, that is, it determines the solid angle $\Omega(D_1 \cup D_2')$ of the union of the near end cap and the projection of the far one. Therefore, $g = 1 - \Omega(D_1 \cup D_2')/2\pi$, as in \cref{eq:eff_end_caps}.
Following~\citet{Lin_2026}, the detector radius is used to normalise the geometric parameters,
\begin{equation}\label{eq:lin_norm}
\bar{\rho}_S = \frac{\rho_S}{R}, \qquad
\bar{L} = \frac{L}{R}, \qquad
\bar{h} = \frac{h}{R}, \qquad
h = L - z_S,
\end{equation}
where $h$ is the distance from the source to the nearer end cap $D_1$ (recall that $z_S \ge 0$, see \cref{section:notation}). In these units the detector radius is unity and the cylinder extends from $-\bar{L}$ to $\bar{L}$.

Depending on whether $D_2'$ is contained in $D_1$ or protrudes beyond it,~\citet{Lin_2026} distinguish two geometric cases. In the second case, the integration is divided into two contributions and requires the auxiliary parameters
\begin{equation}\label{eq:sec_case_params}
\bar{R}_2 = \frac{\bar{h}}{2\bar{L}-\bar{h}},
\qquad
\bar{m} = \frac{2\bar{L}\,\bar{\rho}_S}{2\bar{L}-\bar{h}},
\end{equation}
the normalised radius of $D_2'$ and the normalised distance between the centres of $D_1$ and $D_2'$.

The solid angle is obtained by converting the integral over the area of $D_1 \cup D_2'$ into an integral along its boundary. If $D_2'$ lies entirely inside $D_1$, the boundary is simply the rim of $D_1$. Otherwise it consists of the arc of $D_1$ and the protruding arc of $D_2'$. In both cases $\Omega$ reduces to a one-dimensional integral,
\begin{equation}
  \Omega =
  \begin{cases}
    \displaystyle \int_{0}^{2\pi} F(\theta)\, d\theta ,
      & \bar{m} + \bar{R}_2 \le 1 , \\[2.2ex]
    \displaystyle \int_{\alpha_1}^{2\pi - \alpha_1} F(\theta)\, d\theta
    + \int_{-\pi + \alpha_2}^{\pi - \alpha_2} G(\theta)\, d\theta ,
      & \bar{m} + \bar{R}_2 > 1 ,
  \end{cases}
  \label{eq:lin_cases}
\end{equation}
with the integrands
\begin{equation}
  F(\theta) =
  \frac{\bar{h}\,\sin^{2}\theta}
       {\left[(\cos\theta - \bar{\rho}_S)^{2} + \bar{h}^{2}\right]
        \left[(\cos\theta - \bar{\rho}_S)^{2} + \sin^{2}\theta
              + \bar{h}^{2}\right]^{1/2}} ,
  \label{eq:lin_F}
\end{equation}
\begin{equation}
  G(\theta) =
  \frac{\bar{h}\,\bar{R}_2^{2}\sin^{2}\theta}
       {\left[(\bar{R}_2\cos\theta + \bar{m} - \bar{\rho}_S)^{2}
              + \bar{h}^{2}\right]
        \left[(\bar{R}_2\cos\theta + \bar{m} - \bar{\rho}_S)^{2}
              + \bar{R}_2^{2}\sin^{2}\theta + \bar{h}^{2}\right]^{1/2}} ,
  \label{eq:lin_G}
\end{equation}
where $\theta$ parametrises the rim of the corresponding disc. In the second case, the integration limits are the half-angles at which the two circles intersect. The abscissa of their common chord is
\begin{equation}
  X_c = \frac{1 - \bar{R}_2^{2} + \bar{m}^{2}}{2\bar{m}} ,
  \label{eq:lin_xc}
\end{equation}
from which
\begin{equation}
  \cos\alpha_1 = X_c , \qquad
  \cos\alpha_2 = \frac{\bar{m} - X_c}{\bar{R}_2} .
  \label{eq:lin_alpha}
\end{equation}
The first integral in \cref{eq:lin_cases} then runs over the arc of $D_1$ that lies outside $D_2'$, and the second over the arc of $D_2'$ that protrudes beyond $D_1$. Since $\bar{h} \le \bar{L}$ implies $\bar{R}_2 \le 1$, the second case requires $\bar{m} > 0$, so that \cref{eq:lin_xc} is always well defined.

The implementation used in this work follows the formulation of~\citet{Lin_2026} directly, including the normalisation of the geometric parameters and the distinction between the two geometric cases, with one caveat. We identified a notational inconsistency in the expression for the parameter $\bar{R}_2$ in the cited model (\cref{eq:sec_case_params}) and corrected it accordingly. Originally, the aforementioned parameter was given as:
\begin{equation*}
    \bar{R}_2 = \frac{\bar{L}-\bar{h}}{\bar{L}+\bar{h}}.
\end{equation*}
The two forms agree if, and only if, the symbol $h$ in the original expression is understood as the axial coordinate $z_S$ rather than as the distance from the source to the nearer end cap, the meaning it carries in the accompanying expression for $\bar{m}$. The discrepancy therefore appears to be a notational inconsistency rather than an error in the model itself.

Numerical integration of the resulting expressions was performed using an adaptive quadrature method. This implementation was used solely as an independent reference against which the results of the proposed method were compared. With the corrected expression for $\bar{R}_2$, the~\citet{Lin_2026} model produced results nearly identical to those obtained with \ac{PET-GEOM} for all cases considered in this work. Therefore, the results of the corrected model are, for the most part, not presented separately, as the two methods are effectively equivalent within the scope of the present comparison.

\section{Simplified Monte Carlo sampling}
\label{appendix:simpleMC}

For each simulated event, a random emission direction $d$ was generated by drawing its three Cartesian components from independent standard normal distributions and subsequently normalising the resulting vector,

\begin{equation}
    d
    =
    \frac{\left(d_x,d_y,d_z\right)}{\sqrt{d_x^2+d_y^2+d_z^2}}.
\end{equation}

This procedure produces an isotropic distribution of directions on the unit sphere.
The position of the radioactive source was fixed at $S=(x_S,y_S,z_S)$. The trajectories of the two correlated photons were represented by the same straight line passing through the source point,

\begin{equation}
    l(t)=S+t\,d,
\end{equation}
where the two photons propagate in opposite directions and therefore correspond to positive and negative values of $t$.
For the cylindrical detector, the intersection with the lateral surface was determined analytically by substituting the trajectory equation into the cylinder equation $x^2+y^2=R^2$. This results in the quadratic equation
\begin{equation}
    a_2 t^2+a_1 t+a_0=0,
\end{equation}
with
\begin{align}
    a_2 &= d_x^2+d_y^2,\notag\\
    a_1 &= 2\left(x_Sd_x+y_Sd_y\right),\\
    a_0 &= x_S^2+y_S^2-R^2. \notag
\end{align}

A valid intersection requires a positive discriminant, ensuring that the trajectory intersects the infinite cylindrical surface at two distinct points. The resulting intersection coordinates were subsequently tested against the finite axial extent of the detector. An event was classified as a successful detection if both intersection points satisfied

\begin{equation}
    |z| \leq L
\end{equation}

Thus, trajectories intersecting the lateral surface outside the active axial range were rejected.

\bibliography{cas-refs}

\end{document}